\documentclass[%
 reprint,
showpacs,preprintnumbers,
nofootinbib,%descomentar para usar footnotes
 amsmath,amssymb,
 aps,
prc,
superscriptaddress]{revtex4-1}

\usepackage{graphicx}% Include figure files
\usepackage{dcolumn}% Align table columns on decimal point
\usepackage{bm}% bold math

\usepackage{hyperref}
\usepackage{booktabs}
\usepackage{subfigure}
\usepackage[pdftex]{color}
\usepackage{float}
\usepackage{soul}

\begin{document}

%\preprint{APS/123-QED}

\title{Masquerading hybrid stars with dark matter}

\author{Carline Biesdorf}
\email{carline.fsc@gmail.com}
\affiliation{%
 Departamento de Física, CFM - Universidade Federal de Santa Catarina;  C.P. 476, CEP 88.040-900, Florianópolis, SC, Brasil 
}
\author{Jürgen Schaffner-Bielich}
\affiliation{Institut f\"ur Theoretische Physik, J. W. Goethe Universit\"at, 
  Max von Laue-Str.~1, 60438 Frankfurt am Main, Germany}
\author{Laura Tolos}
\affiliation{
Institute of Space Sciences (ICE, CSIC), Campus UAB, Carrer de Can Magrans, 08193, Barcelona, Spain} 
\affiliation{
Institut d'Estudis Espacials de Catalunya (IEEC), 08860 Castelldefels (Barcelona), Spain}
\affiliation{Frankfurt Institute for Advanced Studies, Ruth-Moufang-Str. 1, 60438, Frankfurt am Main, Germany 
}%

\begin{abstract}

\textbf{Abstract}: We investigate the influence of dark matter on hybrid stars. Using a two-fluid approach, where normal and dark matter components interact only gravitationally, we explore how dark matter can trigger the appearance of quark matter in neutron stars for unprecedented low masses. Our findings reveal that dark matter increases the central pressure of neutron stars, potentially leading to the formation of hybrid stars with quark cores even at very low compact star masses. The critical mass for the appearance of quark matter decreases with increasing dark matter content. We introduce the concept of "masquerading hybrid stars", where dark matter admixed stars exhibit similar mass-radius relations to purely hadronic stars, making it challenging to distinguish between them based solely on these parameters. Additionally, we identify a unique class of objects termed "dark oysters", characterized by a large dark matter halo and a small normal matter core, highlighting the diverse structural possibilities for compact stars influenced by dark matter.

\vspace{0.2cm}

\textbf{Keywords:} dark matter, hybrid stars, two-fluid approach, dark oysters
  
\end{abstract}

%%%%%%%%%%%%%%%%%%%%%%%%%%%%%%%%%%%%%%%%%%%%%%%%%%%%%%%%%%%%%%%%%%%%%%%%%%%%%%%%%

\maketitle

\let\clearpage\relax

\section{Introduction}

Astrophysical and cosmological observations indicate that most of the mass of the Universe appears in the form of non-baryonic mass/energy \cite{Bertone_2005,refId0,refId02}.
%\cite{Bertone_2005,DEL_POPOLO_2014,refId0,refId02}. 
Whereas the Universe is composed of only 4.9$\%$ of baryonic matter, an invisible form of matter called dark matter (DM), whose existence is inferred from its gravitational effects,  rises up to 26.4$\%$ \cite{ade2016planck}. Another component, dark energy (DE), whose existence is related to the accelerated expansion of the Universe, generates 68.7$\%$ of its total mass.

The nature of DM is, however, still puzzling. A lot of candidates have been proposed, with masses ranging from $10^{-33}$~GeV (Fuzzy DM) to $10^{15}$~GeV (Wimpzillas). 
%Among those, WIMPs (Weakly interacting massive particles), axions, and sterile neutrinos, have been favored \cite{deliyergiyev2019dark}.
Whereas there are direct methods for detecting DM using particle accelerators \cite{ATLAS_coll,CMS_coll} or analyzing DM scattering off nuclear targets in terrestrial detectors \cite{KLASEN20151}, no evidence of its existence has been produced so far. An alternative for testing the possible effects of DM are the compact objects (COs), such as white dwarfs (WDs) and neutron stars (NSs). Because of their extreme densities, the probability of the interaction of DM with normal (or ordinary) matter (NM) could be large and the DM capture would be increased \cite{goldman1989weakly,PhysRevLett.109.061301,Kouvaris_2008}.

The authors of \cite{goldman1989weakly,PhysRevLett.107.091301,10.1093/mnrasl/slv049,Acevedo_2021,PhysRevD.107.083012,PhysRevLett.131.091401,Bhattacharya_2024} have investigated the gravitational collapse of NSs caused by DM accretion in various indirect searches for DM, aiming to establish constraints on its properties. Additionally, \cite{PhysRevD.83.083512} examines the accretion of DM in Sun-like and supermassive stars, followed by their collapse into either NSs or WDs. There have also been investigations into how the presence of DM affects the cooling patterns of compact stars, which will ultimately undergo self-annihilation \cite{Kouvaris_2008,PhysRevD.77.043515,PhysRevD.82.063531,PhysRevD.81.083520,de2010neutron,PhysRevD.99.043011,Bhat_cooling}. Furthermore, studies have focused on the alterations in the kinematic properties of NSs resulting from the accretion of self-annihilating DM \cite{PEREZGARCIA20126}.

Based on the idea that the current DM abundance has a similar origin as visible matter, an appealing alternative to WIMPs is the asymmetric dark matter (ADM) model. WIMPs are supersymmetric particles, based on the assumption of a symmetry between bosons and fermions. If, on the other hand, nature is parity symmetric, we have a different form of DM, mirror matter \cite{deliyergiyev2019dark}. Because ADM is non-annihilating it can accumulate, producing changes in mass and radius of the stars, possibly forming extraordinary compact NSs. Comparing the mass-radius relation predicted by star models with NM and with NM admixed with DM in NSs, it is possible to extract information on DM and the equation of state (EoS) of the NSs. Several studies in this regard have been already performed \cite{de2010neutron,blinnikov1983possible,khlopov1989observational,li2012too,sandin2009effects,leung2011dark,leung2012equilibrium,xiang2014effects,Goldman_2013,khlopov2013fundamental,mukhopadhyay2016quark,tolos2015dark,dengler2021erratum,rezaei2017study,panotopoulos2017dark,nelson2019dark,ellis2018dark,PhysRevD.99.083008,deliyergiyev2019dark,DELPOPOLO2020100484,PhysRevD.102.063028,dengler2022second,PhysRevD.105.023001,10.1093/mnras/stab1056,Guha_2021,Miao_2022,10.1093/mnras/stac2675,Ferreira_2023,PhysRevD.109.043029,Hippert_2023,Cassing_2023,astronomy1010005,PhysRevD.106.123027,lenzi2023dark,particles6010012,PhysRevD.108.103016,PhysRevD.109.043030,Sagun_2023,mariani2024constraining,PhysRevD.108.064009,LIU2023101338,BRAMANTE20241,particles7010010,PhysRevD.109.043038,particles7010011,liu2024darkmattereffectsproperties,ciarcelluti2011have,barbat2024comprehensive,pitz2024generating} 
including studies of excitation modes in the presence of DM \cite{Shirke:2023ktu,Shirke:2024ymc}. 
However, only a very few of these kind of analysis have been performed on hybrid stars.
To the best of our knowledge, only refs.~\cite{lenzi2023dark} and \cite{Pal_2024} have performed such studies.

Once the quantum chromodynamics (QCD) phase diagram with its possible phase transitions became a research field of high interest, the possibility that NSs could, in fact, contain both a hadronic and a quark phase, started to be explored \cite{ivanenko1965hypothesis,universe7080267,annala2020evidence}. Asymptotic freedom enables matter to become deconfined when density increases even at low temperatures. Thus, as the density augments toward the star's core, quarks may become more energetically favorable than baryons, leading to the possibility that the core of a NS could be made up of deconfined quarks. If the entire star does not convert itself into a quark star, as suggested by the Bodmer-Witten conjecture \citep{bodmer1971collapsed,witten1984cosmic}, the final composition is a quark core surrounded by an hadronic layer. This is what is generally called a hybrid star~\cite{lukacs1987thermodynamical,glendenning1992first}. We note that in modern terms the QCD phase transition of interest at high densities is the chiral phase transition with the chiral condensate being the order parameter not the deconfinement phase transition. However, of interest for us is that there is the possibility of a first oder phase transition at high densities to be studied below.

Given that DM may compress even further what is already an incredible dense object, it is interesting to analyze how the existence of DM could influence the presence of quark matter (QM) in the interior of NSs. In the present paper we investigate this possible scenario. To that end we consider an EoS for a hybrid star built via Maxwell construction from a Quantum Hadrodynamics (QHD)-based model, for nucleons and hyperons, and a MIT-based model, for u-d-s quarks matter, as already done in~\cite{lopes2022hypermassive}. For the DM EoS we use a non-self annihilating self-interacting Fermi gas with different interaction strengths and two different particle masses, $m_D=5$~GeV and $m_D=100$~GeV~\cite{narain2006compact}. These two fluids, NM and DM, only interact gravitationally. Compared to Refs.~\cite{lenzi2023dark} and \cite{Pal_2024}, even though we use the same model for NM, in these previous works the authors only considered one fluid made of NM interacting with DM through the exchange of a Higgs boson in the former and through the exchange of a scalar as well as a vector meson in the latter. So, to the best of our knowledge, the present work is the first to analyze the effect of DM on hybrid stars using a two-fluid approach.

The present paper is organized as follows. In Section~\ref{sec:TOV} we present the coupled Tolman-Oppenheimer-Volkov (TOV) equations used to obtain the mass-radius diagrams for a two-fluid system and in Section~\ref{sec:EoS} we show the EoSs for these two fluids. 
In Section~\ref{sec:stability} we describe the stability analysis performed in the present work, which follows the recent procedure
described in~\cite{Hippert_2023}, that considers the changes in stability that NM might induce on DM, and vice versa. The analysis of the results starts in Sec.~\ref{sec:results} with Subsection~\ref{sec:DMAHS}, where we show that, in fact, the addition of DM can trigger the early appearance of   QM, thus leading to hybrid stars much earlier than expected due to the presence of DM, an effect we refer to as masquerading hybrid stars with DM.
%and, in not changing much the mass and radius of NSs that otherwise would not be hybrid, 
%the DM may actually be masquerading hybrid stars. 
In Section~\ref{sec:quantity} we discuss the size of the QM core in this hybrid configurations with DM and how it relates with the different interaction strengths and particle mass of the DM. And, finally, in Section~\ref{sec:dark_oysters} we take a closer look to the strongly interacting DM case for a small DM particle mass, demonstrating that compact stars with a small NM core containing   QM can be produced with very large DM radii.
%of 5~GeV, because this DM parametrization yields very different results from all the other parametrizations explored in this work, given rise to 

\section{Stellar Structure Equations}\label{sec:TOV}

In the following we describe the basic equations  of our investigations of DM admixed hybrid stars, where DM is a non-self annihilating self-interacting Fermi gas and hybrid matter is made of NM formed by nucleons, hyperons and   QM. These two fluids only interact gravitationally and, hence, in order to compute the macroscopic properties of DM admixed hybrid stars, we solve the TOV equations, which, in their dimensionless form are given~\cite{narain2006compact}:
\begin{align}\label{eq:TOV}
    & \frac{dp'_{NM}}{dr}=-(p'_{NM}+\epsilon'_{NM}) \frac{d\nu}{dr},\nonumber\\
    & \frac{dm_{NM}}{dr}=4 \pi r^2 \epsilon'_{NM},\nonumber\\
    & \frac{dp'_{DM}}{dr}=-(p'_{DM}+\epsilon'_{DM}) \frac{d\nu}{dr},\nonumber\\
    & \frac{dm_{DM}}{dr}=4 \pi r^2 \epsilon'_{DM},\nonumber\\
    & \frac{d\nu}{dr}=\frac{(m_{NM}+m_{DM})+4\pi r^3(p'_{NM}+p'_{DM})}{r(r-2(m_{NM}+m_{DM}))}     ,
\end{align}
where $p'=P/m^4_{D}$ and $\epsilon'=\epsilon/m^4_{D}$ are the dimensionless pressure and energy density, respectively, being $m_{D}$ the DM particle mass\footnote{In order to obtain the dimensionless quantities for each fluid  we could also divide the physical quantities by a NM particle, as the neutron mass. However, it is important to divide all physical quantities by the same mass scale.}. The physical mass and radius of each species are given by $R_i=(M_p/m^2_{D})r_i$ and $M_i=(M^3_p/m^2_{D})m_i$, respectively, where $i=\{NM,DM\}$ and $M_p$ is the Planck mass~\cite{narain2006compact}. 

As will become clear later on, in order to analyze the stability of the mass-radius configurations, we also need to compute the total number of particles of each species. And for that we solve the following equation for the two number conservations together with the TOV equations above:
\begin{eqnarray}\label{eq:number-TOV}
    \frac{dN'_{i}}{dr}=4 \pi \frac{n'_{i}}{\sqrt{1-2(m_{NM}+m_{DM})/r}}r^2
\end{eqnarray} 
where $n_i'=n_i/m_D^3$ and $N'_i$ are the dimensionless number density and the total number, respectively, of $i=\{NM,DM\}$. We obtain the number of particles for each species by re-scaling them as $N_i=N_i' \cdot M_p^3/m_f^3$ (see Ref.~\cite{narain2006compact}).

\section{Microscopic models}\label{sec:EoS}

\subsection{DM Equation of State}

\begin{figure*}
\begin{centering}
 \includegraphics[angle=0,width=0.47\textwidth]{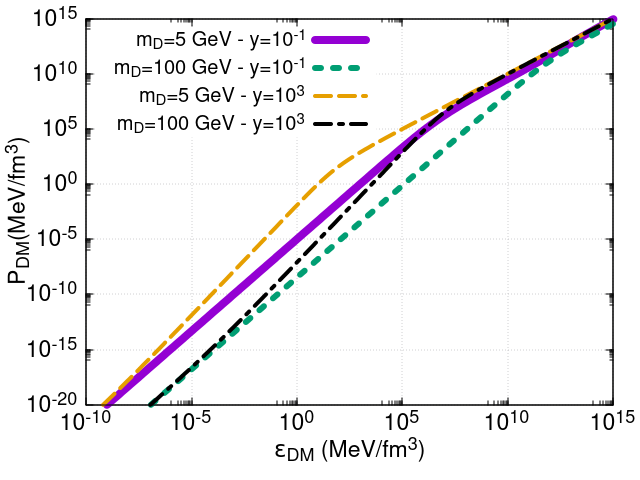}
\caption{DM EoSs for weakly ($y=10^{-1}$) and strongly ($y=10^{3}$) interacting matter and for the two DM particle masses ($m_{D}= 5$~GeV and $m_D=100$~GeV) considered in this work. %{\bf \color{blue} (Laura:I would put different types of lines (dashed, dotted..) apart from the color, because sometimes you do not print in color)} 
}\label{fig:DM_EoS}
\end{centering}
\end{figure*}

For DM the EoS is taken from~\cite{narain2006compact} for a non-self annihilating self-interacting Fermi gas. The dimensionless energy density, pressure and number density are given by, respectively: 
\begin{align}
    \epsilon'_{DM}&=%\frac{\epsilon_{DM}}{m_D^4}=\frac{1}{\pi^2} \int_0^z x^2 \sqrt{1+x^2}dx + \left( \frac{1}{3\pi^2} \right)^2 y^2 z^6 \\ \nonumber
    %&= 
    \frac{1}{8\pi^2} \left[ (2z^3+z)(1+z^2)^{1/2}-{\rm sinh}^{-1}(z)\right]\\ \nonumber
    &+\left( \frac{1}{3 \pi^2} \right)^2 y^2 z^6,\\
    p'_{DM}&=%\frac{P_{DM}}{m_D^4}=\frac{1}{3\pi^2} \int_0^z \frac{x^4}{\sqrt{1+x^2}}dx + \left( \frac{1}{3 \pi^2}\right)^2 y^2 z^6 \\ \nonumber
    %&=
    \frac{1}{24\pi^2} \left[ (2z^3-3z)(1+z^2)^{1/2}+3{\rm sinh}^{-1}(z)\right] \\ \nonumber 
    & + \left( \frac{1}{3 \pi^2} \right)^2 y^2 z^6\\
   \nonumber 
    n'_{DM}&=\frac{n_{DM}}{m_{D}^3}=\frac{z^3}{3\pi^2},
\end{align}
where $z$ is the dimensionless Fermi momentum and $y=m_D/m_I$ %$y=m_f/m_I$ 
the interaction strength, with $m_I$ being the energy scale of the interaction between fermions. Note that for $y\ll 1$ the EoS will be the one of an ideal Fermi gas. Here we will explore the cases of $y$ varying from weakly interacting ($y=10^{-1}$) to strongly interacting ($y=10^{3}$) DM. As for the mass of the DM particle, we will also explore two cases: $m_{D}= 5$~GeV and $m_D=100$~GeV. 
In Fig.~\ref{fig:DM_EoS} we show four representative dimensionful
DM EoSs for the two DM particle masses and for the two different interaction strength parameters discussed here.

\subsection{NM Equations of State}

For the NM we construct a hybrid EoS using the Maxwell construction as criteria to define the hadron-quark phase transition.

To describe hadronic matter we use the Walecka Model~\cite{NLWM} with non-linear terms~\cite{boguta/bodmer}, $\omega$-$\rho$ meson coupling terms and inclusion of the $\phi$ meson that only couples to the hyperons. The EoS can be derived from the following Lagrangian density: 
\begin{widetext}
\begin{center}
\begin{align}\label{eq:lagrang_NLWM}
    \mathcal{L}_{NLWM}& = \sum_{B}\overline{\psi}_B [\gamma_\mu(i\partial^\mu - g_{B \omega}\omega^\mu - g_{B \rho} \frac{\vec{\tau}_B}{2} \vec{\rho}^\mu - g_{B \phi} \phi^\mu)%\nonumber \\
    -m^*_B]\psi_B%\nonumber \\
    +\frac{1}{2}\partial_\mu \sigma \partial^\mu \sigma - \frac{1}{2} m_\sigma^2 \sigma^2 - \frac{1}{3!}  \kappa \sigma^3 - \frac{1}{4!}  \lambda  \sigma^4  \nonumber \\
    & - \frac{1}{4} \Omega^{\mu\nu}\Omega_{\mu\nu} + \frac{1}{2}m_\omega^2 \omega_\mu \omega^\mu %+ \frac{1}{4!}\xi g_{N \omega}^4 (\omega_\mu \omega^\mu)^2 
     - \frac{1}{4} \vec{R}_{\mu \nu} \vec{R}^{\mu \nu} + \frac{1}{2}m_\rho^2 \vec{\rho}_\mu \vec{\rho\,}^\mu + \Lambda_v g_{N \omega}^2 g_{N \rho}^2\omega_\mu \omega^\mu \vec{\rho}_\mu \vec{\rho\,}^\mu
     - \frac{1}{4}\Phi^{\mu \nu} \Phi_{\mu \nu} +\frac{1}{2} m_\phi^2 \phi_\mu \phi^\mu , %\nonumber \\, 
\end{align}
\end{center}
\end{widetext}
where the Dirac spinor $\psi_B$ represents the baryons with the effective mass $m_B^*=m_B-g_{B \sigma} \sigma$, $\vec{\tau}_B$ are the corresponding Pauli matrices, $g_{Bi}$ are the coupling constants of the mesons $i=\sigma, \omega, \rho, \phi$ with the baryon $B$, $m_i$ is the mass of the meson $i$, $\Omega_{\mu \nu}=\partial_\mu \omega_\nu - \partial_\nu \omega_\mu$, $\vec{R}_{\mu \nu}=\partial_\mu\vec{\rho}_\nu - \partial_\nu\vec{\rho}_\mu - g_\rho(\vec{\rho}_\mu \times \vec{\rho}_\nu)$ and $\Phi_{\mu \nu}=\partial_\mu \phi_\nu - \partial_\nu \phi_\mu$. The quantities $\kappa$ and $\lambda$ are scalar self-interaction constants responsible for softening the EoS of symmetric nuclear matter around saturation density, while allowing to obtain a realistic value for the compression modulus of nuclear matter, introduced in \cite{boguta/bodmer, Boguta:1981px}. The 
$\Lambda_v$ is the coupling constant of the mixed quartic isovector-vector interaction that modifies the density dependence of the nuclear symmetry energy \cite{Horowitz:2000xj,Horowitz:2001ya}. The $B$ sum extends over the octet of the lightest baryons $\{n, p, \Lambda, \Sigma^-, \Sigma^0, \Sigma^+, \Xi^-, \Xi^0\}$.

After applying the mean-field approximation, the EoS can be easily obtained from Eq.~(\ref{eq:lagrang_NLWM}). The mesonic fields are
\begin{eqnarray}
%\label{Eq.Movimento-sigma0}
	\sigma_0 &=& \sum_B \frac{g_{B\sigma} n^s_B}{m_\sigma^2} - \frac{\kappa}{2}\frac{\sigma_0^2}{m_\sigma^2} - \frac{\lambda}{6} \frac{\sigma_0^3}{m_\sigma^2}, \\
%\label{Eq.Movimento-omega0}
	\omega_0 &=& \sum_B \frac{g_{B \omega} n_B}{m_\omega^2}  - \frac{\xi}{6} \frac{g_\omega^4 \omega_0^3}{m_\omega^2} - \frac{2\Lambda_v g_\omega^2 g_\rho^2 \overline{\rho}_{0(3)}^2\omega_0}{m_\omega^2}, \\
%\label{Eq.Movimento-rho0}
	\overline{\rho}_{0(3)} &=& \sum_B \frac{g_{B\rho} I_3 n_B}{m_\rho^2} - \frac{2\Lambda_v g_\omega^2 g_\rho^2 \omega_0^2 \overline{\rho}_{0(3)}}{m_\rho^2}, \\
    \phi_0 &=& \sum_B \frac{g_{B\phi}n_B}{m_\phi^2},
\end{eqnarray}
where $n^s_B$ and $n_B$ are the scalar and baryon densities for each baryon species, with the total scalar $n_S$ and baryon (hadron) $n_H$ densities given by
\begin{eqnarray}
    n_S&=&\sum_B \frac{1}{\pi^2} \int dk_B \cdot k_B^2 \frac{m_B^*}{\sqrt{k_B^2+m_B^{*2}}} , \\
%\label{eq:density-hadrons}
    n_H&=&\sum_B \frac{1}{\pi^2} \int dk_B \cdot k_B^2.
\end{eqnarray}
%All sums are over all the baryon octet.
The energy density and pressure then read
\begin{align}\label{eq:energia_NLWM}
	\epsilon_H = & \sum_B \frac{1}{\pi^2} \int dk_B \cdot k_B^2 \sqrt{k_B^2 + m_B^{*2}} \nonumber\\
	& + \frac{1}{2}(m_\sigma^2 \sigma_0^2 + m_\omega^2 \omega_0^2 + m_\rho^2 \overline{\rho}_{0(3)}^2 + m_\phi^2 \phi_0^2) \nonumber\\
	& + \frac{\kappa}{3!} \sigma_0^3 + \frac{\lambda}{4!} \sigma_0^4 + \frac{\xi}{8}g_\omega^4 \omega_0^4  + 3 \Lambda_v g_\omega^2 g_\rho^2 \omega_0^2 \overline{\rho}_{0(3)}^2 ,
\end{align}
\begin{align}
\label{eq:pressao_NLWM}
	P_H = & \sum_B \frac{1}{3\pi^2} \int dk_B \cdot  \frac{k_B^4}{\sqrt{k_B^2 + m_B^{*2}}} \nonumber\\
	& - \frac{1}{2}(m_\sigma^2 \sigma_0^2 - m_\omega^2 \omega_0^2 - m_\rho^2 \overline{\rho}_{0(3)}^2 - m_\phi^2 \phi_0^2) \nonumber\\
	& - \frac{\kappa}{3!} \sigma_0^3 - \frac{\lambda}{4!} \sigma_0^4 + \frac{\xi}{4!}g_\omega^4 \omega_0^4 + \Lambda_v g_\omega^2 g_\rho^2 \omega_0^2 \overline{\rho} _{0(3)}^2 .
\end{align}
We choose the NL3$^* \omega \rho$ model, which is the NL3$^*$ parametrization proposed in \cite{lalazissis2009effective} with the addition of the $\omega \rho$-channel as done in \cite{lopes2022nature}. This parametrization has the following nuclear saturation parameters: $n_0=0.150$~fm$^{-3}$, $E/A=16.3$~MeV, $K=258$~MeV,  $E_{sym}=30.7$~MeV, $L=42$~MeV and $M^{*}/M=0.59$, which satisfy the phenomenological constraints taken from \cite{dutra2014relativistic,oertel2017equations}. This parametrization 
%This parametrization satisfies the nuclear matter properties at saturation density \cite{dutra2014relativistic,oertel2017equations} and
also reproduces maximum star masses above $2$~M$_\odot$, even when hyperons are included. The main parameters %of these parametrizations %as well as the main nuclear properties 
are presented in Table~\ref{tab:parameters}.

%\begin{center}
%\begin{table}%[ht]
%\begin{center}
%\caption{Parameters of the model utilized in this work and their prediction for the symmetric nuclear matter properties at the saturation density. The parametrization is taken from ref.~\cite{biesdorf2023qcd} and the phenomenological constraints are taken from \cite{dutra2014relativistic} and ~\cite{oertel2017equations}. The meson masses $m_\sigma$, $m_\omega$ and $m_\rho$ as well as $\kappa$ are given in MeV. The nucleon and $\phi$ masses are fixed at $M=939$~MeV and $m_\sigma=1020$~MeV, respectively. {\color{blue} \bf (Laura: you put too many digits. Could you reduce it to three significant digits upmost? And maybe separate parameters from constraints. Two tables?)}}
%\label{tab:parameters}
%\scalebox{0.90}{
%\begin{tabular}{|c|c||c|c|c||c|}
%\hline 
%  & Parameters & &  Constraints  & NL3$^*\omega \rho$  \\
% \hline
% $m_\sigma$ & 502.574  &$n_0$(${\rm fm^{-3}}$) & 0.148 - 0.170 & 0.150 \\
% \hline
%  $m_\omega$ & 782.600 & $E/A$ (MeV) & 15.8 - 16.5 & 16.3 \\
%  \hline
%  $m_\rho$ & 763.00 & $K$ (MeV)& 220 - 260 & 258 \\
% \hline
%$g_{N\sigma}$ & 10.0944 & $E_{sym}$ (MeV) & 28.6 - 34.4 & 30.7  \\
% \hline
% $g_{N\omega}$ &  12.8065 & $L$ (MeV) & 36 - 86.8 & 42\\
% \hline 
%$g_{N\rho}$ &  14.4410 & $M^{*}/M$ & 0.6 - 0.8 & 0.59  \\
%\hline 
%$\kappa/g_{N\sigma}^3$ & 4.1473 &  &  &   \\
%\hline
%$\lambda/g_{N\sigma}^4$ &  -0.017422 &  & & \\
%\hline 
%$\Lambda_v$ &  0.045 & & &   \\ 
%\hline
%\end{tabular}}
%\end{center}
%\end{table}
%\end{center}

%\begin{widetext}
%\begin{center}
\begin{table*}
\caption{Parameters of the model NL3$^*\omega \rho$ utilized in this work. The parametrization is taken from Ref.~\cite{biesdorf2023qcd}. The meson masses $m_\sigma$, $m_\omega$ and $m_\rho$ as well as $\kappa$ are given in MeV. The nucleon and $\phi$ masses are fixed at $M=939$~MeV and $m_\sigma=1020$~MeV, respectively.}
%\begin{tabular}{@{}ccccccccccc@{}}
\begin{tabular}{p{1.5cm}p{1.2cm}p{1.2cm}p{1.2cm}p{1.5cm}p{1.5cm}p{1.5cm}p{1.9cm}p{2.2cm}p{1.2cm}}
\toprule[0.6pt]
 $m_\sigma$  & $m_\omega$  & $m_\rho$  & $g_{N \sigma}$  & $g_{N \omega}$  & $g_{N \rho}$  & $\kappa/g_{N\sigma}^3$  & $\lambda/g_{N\sigma}^4$  & $\Lambda_v$  \\ \midrule[1.5pt] 
 $502.574$  & $782.600$  & $763.000$  & $10.0944$  & $12.8065$  & $14.4410$  & $4.1473$  & $-0.017422$  & $0.045$ \\ \bottomrule[0.6pt]
\end{tabular}\label{tab:parameters}
\end{table*}
%\end{center}
%\end{widetext}

As for hyperons, we consider the hyperon masses to be $m_\Lambda=1116$~MeV, $m_{\Sigma}=1193$~MeV and $m_{\Xi}=1318$~MeV. The couplings of the hyperons to the vector mesons are related to the nucleon couplings, $g_{N \omega}$ and $g_{N \rho}$, by assuming SU(6)-flavour symmetry, according to the ratios \cite{dover1984hyperon,schaffner1996hyperon,banik2014new,tolos2017equation,Tolos:2016hhl,lopes2014hypernuclear}:
\begin{align}
    &g_{\Lambda \omega} : g_{\Sigma \omega} : g_{\Xi \omega} : g_{N \omega} = \frac{2}{3} : \frac{2}{3} : \frac{1}{3} : 1, \nonumber \\
    &g_{\Lambda \rho} : g_{\Sigma \rho} : g_{\Xi \rho} : g_{N \rho} = 0:2:1:1,\\
    &g_{\Lambda \phi} : g_{\Sigma \phi} : g_{\Xi \phi} : g_{N \omega} = -\frac{\sqrt{2}}{3} : -\frac{\sqrt{2}}{3} : -\frac{2\sqrt{2}}{3} : 1 \nonumber,
\end{align}
noting that $g_{N \phi}=0$. The coupling of each hyperon to the $\sigma$ field is adjusted to reproduce the hyperon potential in symmetric nuclear matter (SNM) derived from hypernuclear observables. We fix this potentials as $U_\Lambda(n_0)=-28$~MeV, $U_\Sigma(n_0)=+30$~MeV and $U_\Xi(n_0)=-4$~MeV and obtain the coupling constants presented in Table~\ref{tab:acoplamentos}.

\begin{table}[]
\caption{Hyperon-$\sigma$ coupling constants adjusted to reproduce the hyperon potential in SNM derived from hypernuclear observables.}
%\begin{tabular}{@{}ccccccccccc@{}}
\begin{tabular}{p{1.5cm}p{1.4cm}p{1.4cm}p{1.4cm}}
\toprule[0.6pt]
 Model  & $g_{\Lambda \sigma}/g_{N \sigma}$  & $g_{\Sigma \sigma}/g_{N \sigma}$  & $g_{\Xi \sigma}/g_{N \sigma}$  \\ \midrule[1.5pt]
 NL3$^*\omega \rho$ & $0.613$  & $0.461$  & $0.279$  \\ \bottomrule[0.6pt]
\end{tabular}\label{tab:acoplamentos}
\end{table}

%As for the couplings of the hyperons to the vector mesons, we assume SU(6)-flavour symmetry, as done in~\cite{biesdorf2023qcd}. This and all other details about the NL3$^* \omega \rho$ parametrization can be found in~\cite{biesdorf2023qcd} and references therein. 

%The detailed calculations of the EoS for symmetric nuclear matter, as well as for $\beta$ stable matter in the QHD formalism are well documented and can be easily found in the literature~\citep{glendenning2000compact,serot1992quantum,universe7080267}.

To describe the quark matter we use the modified MIT Bag model, as introduced in~\cite{lopes2021modified}, with the inclusion of a vector field and a self-interacting vector field. The Lagrangian density is as following: 
%The EoS can be derived from the following Lagrangian density:
\begin{widetext}
\begin{center}
\begin{align}\label{mit}
    \mathcal{L}_{MIT}&=\sum_q \Big\{ \overline{\psi}_q \Big[\gamma^\mu(i \partial_\mu -g_{qqV} V_\mu) - m_q\Big]\psi_q + \frac{1}{2}m_V^2 V_\mu V^\mu + b_4\frac{(g_{uuV}^2 V_\mu V^\mu)^2}{4} - B\Big\}\Theta(\overline{\psi}_q \psi_q) - \frac{1}{2}\overline{\psi}_q \psi_q \delta_S,
\end{align}
\end{center}
\end{widetext}
where the Dirac spinor $\psi_q$ represents the quark with mass $m_q$ running over $u, d$ and $s$, whose values are 4 MeV, 4 MeV and 95 MeV, respectively~\cite{tanabashi2018review}, $g_{qqV}$ the coupling constant, and $m_V$ the mass of the meson. The quantity $\Theta(\overline{\psi}_q \psi_q)$ is a Heaviside function that ensures that the quarks are confined inside the bag and $\delta_S$ is a Dirac function that guarantees continuity of the fields of the quarks on the surface of the bag. Using the mean-field approximation and solving the Euler-Lagrange equations of motion, we obtain the energy eigenvalues for the quarks and for the $V$ field as follows:
\begin{eqnarray}
&& E_q = \mu = \sqrt{m_q^2 + k^2} + g_{qqV}V_0,\\
 &&g_{uuV}V_0  + \bigg ( \frac{g_{uuV}}{m_v} \bigg)^2  b_4 (g_{uuV}V_0)^3 \nonumber \\
 &=& \bigg (\frac{g_{uuV}}{m_v} \bigg ) \sum_{q=u,d,s} \bigg (\frac{g_{qqV}}{m_v} \bigg )n_q . 
%\end{align}
\end{eqnarray}
The baryon density in the quark phase, energy density and pressure are given by
\begin{align}
    n_{Q}&= \sum_q \frac{n_q}{3} = \sum_q \frac{1}{\pi^2} \int dk_q \cdot k_q^2, \label{eq:dens__MIT_auto-interac_T}\\
\epsilon_Q &=  \sum_q \left( \frac{3}{\pi^2}\int dk_q \cdot k_q^2\sqrt{m_q^2 + k_q^2}  \right) + \frac{1}{2}m_V^2V_0^2 + \nonumber \\ 
&+ \frac{3}{4} b_4 (g_{uuV}V_0)^4 + B, \label{eq:energia_MIT_auto-interac_T}\\
    P_Q &=\sum_q \left( \frac{1}{\pi^2}\int dk_q \frac{k_q^4}{\sqrt{m_q^2 + k_q^2}} \right) + \frac{1}{2}m_V^2V_0^2 + \nonumber \\    
    &+ \frac{1}{4} b_4 (g_{uuV}V_0)^4 - B. \label{eq:pressao_MIT_auto-interac_T}
\end{align}
Here we choose the parametrization $B^{1/4}=155$~MeV, for the Bag pressure value, $G_V=(g_{uuV}/m_V)^2=1.0$~fm$^2$ and $b_4=1.2$. As for the relation between the coupling constants we opt to use the ones obtained via symmetry relations, where one has $g_{ssV}=\frac{2}{5} g_{uuV}=\frac{2}{5} g_{ddV}$. This parametrization describes unstable strange matter, % i.e., the parameters are outside the stability window, 
which allows the existence of a quark core that will not convert the whole star into a strange star~\cite{lopes2022hypermassive}.

Neutron stars are charged neutral objects in $\beta$-equilibrium. Therefore, in order to produce $\beta$-stable matter with zero net charge, we also add leptons as a free Fermi gas to both hadron and quark matter and impose the conditions of $\beta$-equilibrium and charge neutrality. For hadron matter,
\begin{eqnarray}
    \mu_B=\mu_n-q_B ~\mu_e \quad \textrm{and} \quad \mu_{e}=\mu_{\mu}, \nonumber \\
    n_p + n_{\Sigma^+} = n_{e^-} + n_{\mu^-} + n_{\Sigma^-} + n_{\Xi^-} ,
\end{eqnarray}
whereas for quark matter
\begin{eqnarray}\label{eq:equil.quim_netr.carga}
  \mu_s =\mu_d = \mu_u + \mu_e  \quad \mbox{and} \quad \mu_e = \mu_\mu , \nonumber \\
  n_{e^-} + n_{\mu^-} = \frac{1}{3}(2n_u -n_d - n_s).
\end{eqnarray}
As for the inner and outer crust of the stars, we use~\cite{negele1973neutron} and~\cite{ruster2006outer}, respectively and for $\epsilon < 3.3 \times 10^3$~g/cm$^3$ we use the Harrison-Wheeler EoS~\cite{harrison1965gravitation}.

We also impose an upper limit on the NM EoS and allow pressures only up to 1000 ${\rm MeV/fm^3}$, which corresponds to a density around $12n_0$, where $n_0$ is the nuclear saturation density. So we stay far below the density at which the EoS from perturbative QCD (pQCD) is known, which is about $40n_s$~\cite{Komoltsev_2022}. Higher pressures would eventually allow for more mass-radius solutions in the NS branch to be obtained for the same amount of DM, as discussed later. However, since our general conclusions are independent of the upper limit of the NM EoS and this study does not account for pQCD effects, we consider the chosen pressure limit to be suitable for the scope of this work.

Note that the EoSs for the hadron and quark phases are dimensionful. Therefore, in order to solve Eqs.~(\ref{eq:TOV}) together with Eq.~(\ref{eq:number-TOV}), we will scale the pressure and energy density with the DM particle mass, as done for the DM EoS.

\subsubsection{The Hybrid EoS}

Once the EoSs for the hadron and quark phases are known, we can build the EoS to describe hybrid stars, i.e., compact stars with a core of QM surrounded by hadron matter. For that purpose we use the Maxwell construction. % In the so-called Maxwell construction, the hadron and quark phases are spatially separated and there is no mixed-phase. The Maxwell construction is not the only possible way to study quark-hadron phase transition. Another one is the so-called Gibbs condition, where there is a mixed phase and the quarks and hadrons can coexist. Some authors~\citep{Maruyama2007,Paoli2010} performed studies on hybrid stars with both, Maxwell and Gibbs construction and concluded that there is no significant difference on the macroscopic properties of the hybrid stars. On the other hand, some studies~\citep{Maslov,Yasutake} point out that the presence of a mixed phase can affect some neutron star properties, especially those  with intermediate masses, although the value of the maximum mass is hardly affected. Moreover, a recent work~\citep{Lugones2021} suggests that the Maxwell construction is favoured if the surface tension between quarks and hadrons is high enough, therefore, we use this construction. 
In this case the necessary conditions for thermodynamic equilibrium of the hadronic and quark phases allowing a first order phase transition are given by:
\begin{equation}
 \mu_0^H = \mu_0^Q \quad \mbox{and} \quad P_0^H = P_0^Q,
\end{equation}
where $\mu_0$ for the hadrons and quarks are
\begin{eqnarray}
 \mu_0^H = \frac{(\sum_B \mu_Bn_B + \sum_l\mu_l^Hn_l^H)}{\sum_B n_B} ,\nonumber \\
 \mu_0^Q = \frac{3(\sum_q \mu_q n_q +\sum_l \mu_l^Q n_l^Q)}{\sum_q n_q} . \label{eq:mu}
\end{eqnarray}
There is no experimental evidence about the value of the baryon chemical potential at the hadron-quark interface, $\mu_0$, at zero temperature. An inferior limit of 1050 MeV was pointed out in~\cite{fukushima2010phase} using the Polyakov loop formalism. In~\cite{lopes2022hypermassive} an upper limit of $\mu_0~=$ 1400 MeV was adopted based on the discussion of Ref.~\cite{annala2020evidence}, where the authors also point out that QM inside massive neutron stars is not only possible but probable. Here, however, we relax this condition and use a higher value for $\mu_0$, as can be seen in Table~\ref{tab:phase_trans_quant}.
As the first hyperons start to appear around $n_H=0.31$~fm$^{-3}$, at a chemical potential of $\mu_H$ = 1135 MeV, which is below the value of $\mu_0$, this choice of parametrizations also justifies the inclusion of these more massive baryons, besides the nucleons.
The quantities shown in table~\ref{tab:phase_trans_quant} for the phase transition point will remain the same throughout this entire work and will play an important role determining at which mass of the NS the hadron-quark phase transition will occur.

\begin{table}
    \centering
    \begin{tabular}{|c|c|c|c|}
    \hline
        $P_0$ (MeV/fm$^3$) & $\mu_0$~(MeV) & $\epsilon_H$ (MeV/fm$^3$) & $\epsilon_Q$ (MeV/fm$^3$) \\
        \hline
        361 & 1698 & 1003 & 1280 \\
         \hline
    \end{tabular}
    \caption{Main thermodynamic values at the hadron-quark phase transition. These quantities will remain the same throughout the entire paper. }
    %The pressure $P_0$ and energy densities $\epsilon$ are given in MeV/fm$^3$.
    \label{tab:phase_trans_quant}
\end{table}
 
\begin{figure}[ht]
\begin{tabular}{cc}
\includegraphics[width=0.47\textwidth,angle=0]{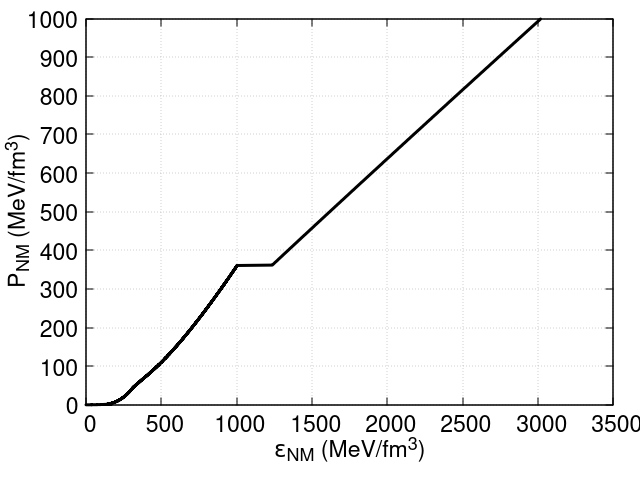} \\%&
\includegraphics[width=0.47\textwidth,angle=0]{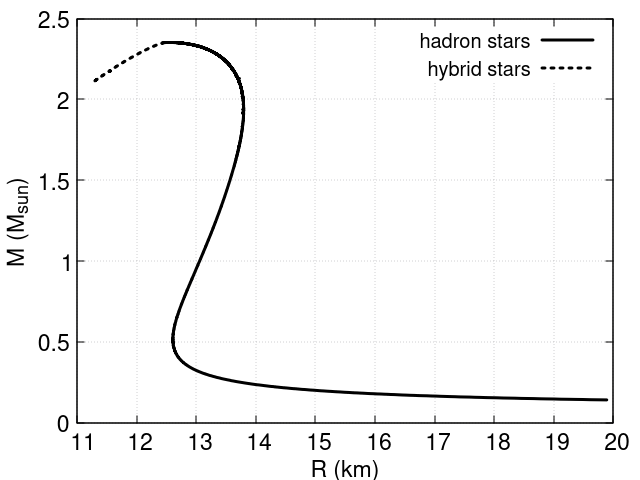} \\
\end{tabular}
\caption{EoS (top) and mass-radius relation (bottom). No stable hybrid star is possible within these parametrizations.} \label{fig:MxR-noDM}
\end{figure}

In order to construct the EoS to describe hybrid stars, one needs to ensure that $uds$-quark matter is unstable ($E/A > 930$ MeV). Otherwise, as soon as the core of the star converts to the quark phase, the entire star may convert into a quark star in a finite amount of time~\cite{olinto1987conversion}. As already mentioned above, the parametrization chosen here ensures that $uds$-quark matter is unstable. However, this specific choice of parameters for the MIT model in combination with the QHD model has an even more profound meaning. If used to describe single-fluid stars, the hadron-quark phase transition happens only when NSs become unstable,
%the critical mass $M_{crit}$ corresponds to the first dynamically unstable star after the maximum mass star,
i.e., with this combination of parametrizations, all the dynamically stable stars ($\partial M/\partial \epsilon_C > 0$) are purely hadronic. The reason for this choice is to emphasise the effect of the DM on the hadron-quark phase transition, as will become clear later on.
Furthermore, the choice for the QHD parametrization ensures that we still meet the constraints imposed by PSR J070+6620~\cite{riley2021nicer} and PSR J0952–0607~\cite{romani2022psr}. In Fig.~\ref{fig:MxR-noDM} we show the hybrid EoS (top) and the corresponding mass-radius diagram (bottom) obtained by solving the single-fluid TOV equations.

%\begin{center}
%\end{center}

\section{Stability analysis}\label{sec:stability}

In order to draw conclusions on the effect of DM on  hybrid stars we have to first check  where our mass-radius configurations are dynamically stable. In the case of a single fluid, the stability analysis is based on considering small perturbations from the hydrostatic equilibrium and then solve a Sturm-Liouville problem, which results in $n$ eigenfrequencies $\omega_n$ that obey the discrete hierarchy $\omega_n^2 < \omega_{n+1}^2$, $n=0,1,2,...$, with $\omega_0^2 > 0$. If $\omega_0^2 < 0$, then the lowest energy mode is imaginary, indicating an instability. To determine the sign of the mode one can, instead of solving the eigenvalue problem, analyze the dependence of the star's total mass and radius as a  function of its central energy density (or central pressure). The extrema in the mass versus energy density (or pressure) indicates a change in the sign of eigenfrequency associated to a certain mode. If the derivative of the radius versus the energy density (or pressure) at that energy density is negative (positive) an even (odd) mode is changing sign. So, starting from low energy densities where all modes are positive, it is then possible to perform the stability analysis for higher energy densities studying the change in sign of the eigenfrequency modes and checking when the lowest one becomes negative (see Ref.~\cite{schaffner2020compact} for more details). 

In Refs.~\cite{tolos2015dark,dengler2021erratum,dengler2022second} the above described stability analysis was applied to a two fluid system. In that case this method is used for the two species of matter separately (NM and DM) and the compact object is found stable only when both species of matter are stable. However, this naive stability analysis does not consider the changes in stability that NM might induce on DM and vice versa. Therefore here we will make use of a different method based on the one developed by \cite{henriques1990stability} and expanded in \cite{Goldman_2013,Kain_2021,di2022fermion,Valdez_Alvarado_2013,nyhan2022dynamical}. More precisely, we will follow closely what was done in \cite{Hippert_2023}. This method is based on the fact that at the onset of unstable radial modes, the total number of fluid elements $N$ must be stationary under the variations of the central energy density $\epsilon_c$, i.e., $\partial N/ \partial \epsilon_c=0$. This criteria is consistent with the more commonly used criteria $\partial M/ \partial \epsilon_c=0$ as, under the assumption of uniform entropy per baryon, the stellar mass is stationary under any transformation $\epsilon(r) \rightarrow \epsilon(r)+\delta \epsilon(r)$ that leaves the total number of fluid elements $N$ unchanged \cite{weinberg1972gravitation} and this is true if and only if the TOV equations are satisfied, i.e., only for equilibrium configurations.

%{\color{blue} The aforementioned criteria was generalized for the case of a two fluid star in \cite{Goldman_2013} {\bf (Laura: here I am puzzled. I thought the previous references you mentioned also did two-fluids. Is not that the case?)}}.In this case 

In the case of a two-fluid star the TOV Eqs.~(\ref{eq:TOV}) and (\ref{eq:number-TOV}) are satisfied if and only if any transformation $\epsilon_i(r) \rightarrow \epsilon_i(r)+\delta \epsilon_i(r)$ that leaves $\delta N_i=0$ also leaves the total mass unchanged $\delta M =0$, with $i=\{NM,DM\}$. 
 
The condition that $N_i$ are stationary % under which $M$ is stationary as well 
is given by
\begin{align}
\label{eq:matrix_onset_instability}
    \begin{pmatrix} 
    \delta N_{NM} \\ \delta N_{DM}    
    \end{pmatrix} 
    &= 
    \begin{pmatrix}
        \partial N_{NM}/ \partial \epsilon_c^{NM} & \partial N_{NM}/ \partial \epsilon_c^{DM} \\
        \partial N_{DM}/ \partial \epsilon_c^{NM} & \partial N_{DM}/ \partial \epsilon_c^{DM}
    \end{pmatrix}
    \begin{pmatrix} 
    \delta \epsilon_c^{NM} \\ \delta \epsilon_c^{DM}    
    \end{pmatrix} \nonumber\\
    &= 0 \,.
\end{align}
As the small shifts in the energy density are not zero, $\delta \epsilon_c^{i} \neq 0$, we have that
\begin{equation}
    \frac{\partial N_{NM}}{\partial \epsilon_c^{NM}} \frac{\partial N_{DM}}{\partial \epsilon_c^{DM}} - \frac{\partial N_{NM}}{\partial \epsilon_c^{DM}} \frac{\partial N_{DM}}{\partial \epsilon_c^{NM}} = 0,
\end{equation}
which is the criteria for the onset of radial instability for two-fluid stars. So we have that, at the onset of instability, $\delta N_i=0$ and $\delta M =0$ under variations of the central energy densities, ($\epsilon_c^{NM},\epsilon_c^{DM}$)$\rightarrow$($\epsilon_c^{NM}+\delta \epsilon_c^{NM},\epsilon_c^{DM}+\delta \epsilon_c^{DM}$).

For a two-fluid system, the matrix above can be diagonalized and one obtains two independent sets of variables, ($\epsilon_C^A, N_A$) and ($\epsilon_C^B, N_B$) corresponding to eigenvalues $\kappa_A$ and $\kappa_B$. As $N_A$ and $N_B$ are linear combinations of $N_{NM}$ and $N_{DM}$, they are also conserved and kept fixed when the star is perturbed. So, if small and independent changes to $\epsilon_C^A$ and $\epsilon_C^B$ are performed, the stable solutions should satisfy:
\begin{equation}\label{eq:eigen_stable}
    \kappa_A > 0 \qquad \mathrm{and} \qquad \kappa_B > 0.
\end{equation}
This generalizes the widely used stability condition $\partial M/ \partial \epsilon_C > 0$ to multi-fluid stars \cite{Hippert_2023}. In the present work we will use the condition given by Eq.~(\ref{eq:eigen_stable}) to determine if our solutions are stable or not.

\section{Results}\label{sec:results}

\subsection{Dark Matter Admixed Hybrid Stars}\label{sec:DMAHS}

Several works have already analyzed the effect of DM on the whole mass-radius diagram, i.e., on the NS branch as well as on the WD branch. Even though different EoSs have been used to describe NM, the results yield the same general conclusions, the main one being that, when one of the central pressures significantly exceeds the other, the fluid with the higher central pressure takes over, causing the system to act as if it consists of a single fluid. Also, the pressure at which DM starts to prevail depends on its particle mass as well as on its interaction strength. For more details the interested reader can refer, for example, to \cite{tolos2015dark,dengler2021erratum,barbat2024comprehensive,dengler2022second}. 
%although this last one uses a slightly different definition for the pressure ratios. 
Here we focus on the NS branch and analyze how the presence of DM could favor the appearance of hybrid stars.

We start by defining a critical mass $M_{\rm crit}$ as the minimum mass of a star that has a quark core, i.e., stars with masses below $M_{\rm crit}$ are purely hadronic or are DM admixed hadronic stars, but without quark matter in its core. At the microscopic level this means that the star with $M_{\rm crit}$ is the one with a central pressure of NM equal to the pressure where the hadron-quark phase transition happens, i.e., $P_0=361$~MeV/fm$^3$, so that all the stars with higher NM central pressure are hybrid stars.

\begin{figure*}[ht]
\begin{centering}
\begin{tabular}{cc}
\includegraphics[angle=0,width=0.47\textwidth]{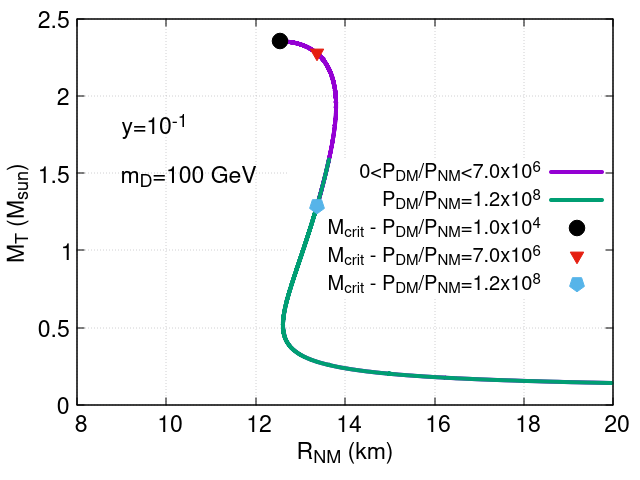} &
\includegraphics[angle=0,width=0.47\textwidth]{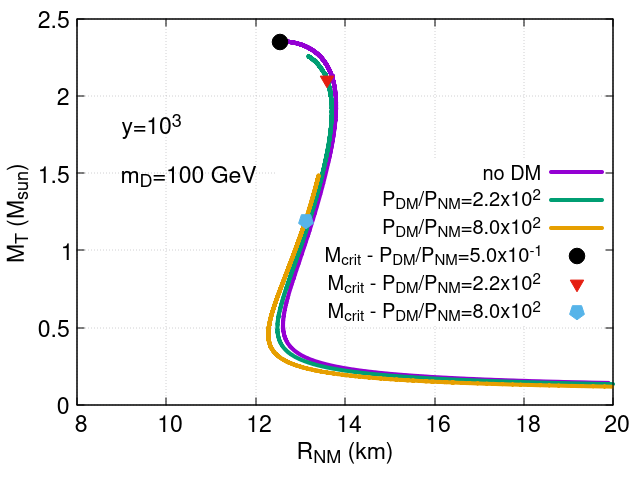} \\
\includegraphics[angle=0,width=0.47\textwidth]{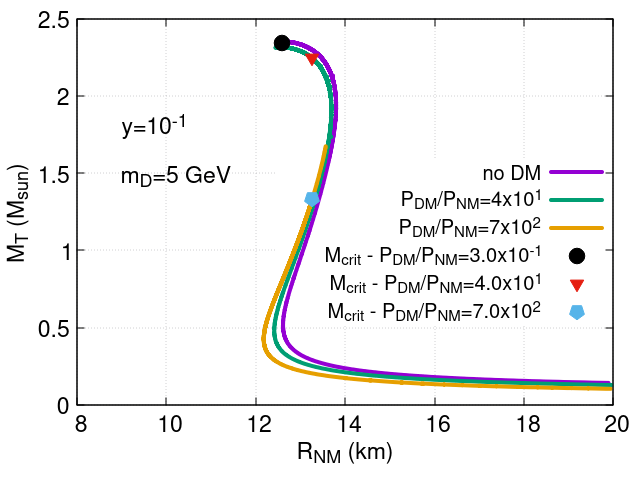} &
\includegraphics[angle=0,width=0.47\textwidth]{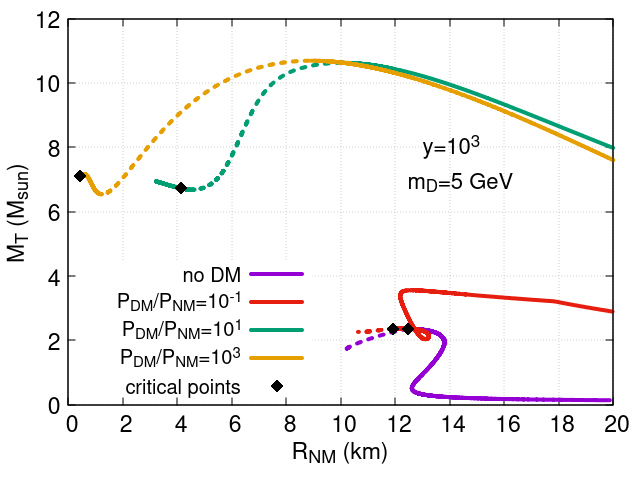} \\
\end{tabular}
\caption{Total mass ($M_T=M_{NM}$+$M_{DM}$) as a function of the observable radius ($R_{NM}$ radius) of DM admixed NSs for different pressure ratios ($P_{DM}/P_{NM}$). We consider weakly $y=10^{-1}$ (left panels) and strongly $y=10^{3}$ (right panels) interacting DM, as well as $m_D=100$~GeV (top panels) and $m_D=5$~GeV (bottom panels). The dots indicate the critical mass $M_{\rm crit}$ at which hybrid stars start to appear for each ratio. The round black dots indicate the first hybrid star to appear for each $y$ and $m_D$. In the bottom right figure we also include unstable results indicated by dashed lines.}
\label{fig:NS_branch}
\end{centering}
\end{figure*}

In Fig.~\ref{fig:NS_branch} we show the total mass ($M_T=M_{NM}$+$M_{DM}$) versus the observable radius ($R_{NM}$) for different ratios of DM central pressure versus the NM  one ($P_{DM}/P_{NM}$). 
We display the stable mass-radius configurations with DM with different colored solid lines, whereas in violet solid lines we depict the mass-radius relation without DM for comparison. Note that in the top left plot we can vary $P_{DM}/P_{NM}$ up to $7 \times 10^6$ and the mass-radius solutions will still be on top of the mass-radius configurations without DM.
%{\bf (Laura: in the left top panel the violet line includes DM, right? But in the other plots the violet line is no DM. Could you change the color as in the other plots to green or yellow or red? And also add the no DM case for comparison, as in the other plots).}  
We consider weakly interacting DM (left panels) and strongly interacting DM (right panels) as well as two values of DM particle masses ($m_D=100$~GeV for top plots and $m_D=5$~GeV for bottom ones). The triangular and hexagonal points in each plot show the location of the $M_{\rm crit}$ for different $P_{DM}/P_{NM}$ ratios. The round black dots indicate the $M_{\rm crit}$ corresponding to the first stable hybrid star to appear, i.e., they  correspond to the minimum value of $P_{DM}/P_{NM}$ necessary for the appearance of a stable hybrid star. 
%{\bf (Laura: the black dot is only for no DM configurations? In the top left panel is shown for cases with DM?)}}
We remind the reader that, without DM, this branch would correspond to purely hadronic stable stars. However, as can be seen, once we add a certain amount of DM, stable hybrid stars start to appear.
We can also observe that the results corresponding to strongly interacting DM with DM particle mass of 5~GeV (right bottom plot) differ substantially from the others. In this last case, we show stable (solid) and unstable (dashed) mass-radius solutions. The black squared dots merely show the location of the $M_{\rm crit}$ for each ratio shown in the plot. For the case without DM and also for $P_{DM}/P_{NM}=10^3$ the configurations for $M_{\rm crit}$ are unstable.
%We will analyze these configurations in more detail later on, in Sec.~\ref{sec:dark_oysters}. 
%Thus, from now on we concentrate on the results for $m_D=100$~GeV as well as $m_D=5$~GeV for weakly interacting DM.
%{\bf (Laura: those are two different cases, right?. For no DM the solution seems stable, because it is the last point in the stable branch, no?)} 

For $m_D=100$~GeV as well as $m_D=5$~GeV with $y=10^{-1}$, we observe in Fig.~\ref{fig:NS_branch} that, as $P_{DM}/P_{NM}$ increases, the stable mass-radius configurations are reduced as compared to the case without DM. This is due to the fact that, as already mentioned above, with the increase of $P_{DM}/P_{NM}$, DM starts to slowly dominate over NM, inducing a change of stability of the configurations. Indeed, the mass-radius solutions migrate to smaller radii and masses as compared to the case without DM, eventually forming dark compact planets (DCPs), with Earth-like or Jupiter-like masses and radii of about one meter or one kilometer, respectively, as shown in Refs.~\cite{tolos2015dark,dengler2021erratum}. The dominance of DM over NM means that the DM mass fraction is greater than one, i.e., $M_{DM}/M_{NM}>1$. In Fig.~\ref{fig:NS_branch} we show only the results where $M_{DM}/M_{NM}<1$, except for the bottom right panel. For the pressure ratios where the maximum mass is no longer reached, the DM already dominates, but only at low NM pressures.
%In Fig.~\ref{fig:NS_branch}}, however, we only see the beginning of this migration because we are only looking at the region where 8 $\leq$~R$_{NM}$ $\leq$~20 km (for most of the cases) and because we increase the DM pressure ratio only up to values where we can see the effect we are searching for, which is the appearance of stable hybrid stars in this region where, without DM, it would not be possible.
For the case of $m_D=5$~GeV and $y=10^3$ we find, however, a very peculiar behavior, with the total mass growing and the NM radius decreasing (while the DM radius increases) with larger $P_{DM}/P_{NM}$ ratios. In this case we also include the results for smaller radii (R$_{NM}$ $\leq$ 8 km). We will analyze these configurations in more details in Sec.~\ref{sec:dark_oysters}. Thus, from now on we concentrate on the results for $m_D=100$~GeV for $y=10^{-1}$ and $y=10^3$ as well as $m_D=5$~GeV for $y=10^{-1}$.

We can now compare the behavior of the mass-radius configurations among these three cases. We find that for weakly interacting DM one needs to add more DM to obtain the first stable hybrid star than for the strongly interacting case, as can be seen if we compare the  results for $m_D=100$~GeV on the top plots of  Fig.~\ref{fig:NS_branch}. For $y=10^{-1}$ the first stable hybrid star appears at $P_{DM}/P_{NM}=1.0 \times 10^4$, whereas for $y=10^{3}$ an hybrid star already happens at $P_{DM}/P_{NM}=5.0 \times 10^{-1}$. Also, for the same interaction strength $y$, one needs to add more DM when the DM particle mass is larger, as can be seen if we compare the results on the left plots of the same figure for $y=10^{-1}$.  For $m_D=100$~GeV the first stable hybrid star appears at $P_{DM}/P_{NM}=1.0 \times 10^4$, whereas for $m_D=5$~GeV it already appears at $P_{DM}/P_{NM}=3.0 \times 10^{-1}$. 

\begin{figure}[ht]
\includegraphics[angle=0,width=0.47\textwidth]{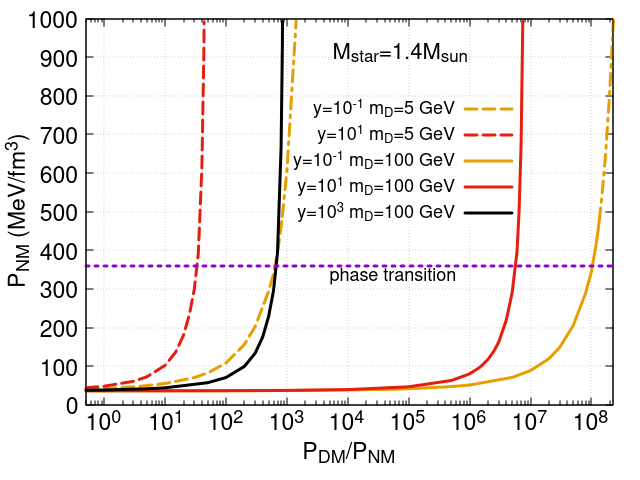} 
\caption{Central pressure of normal matter ($P_{NM}$) as a  function of the ratio of  pressures ($P_{DM}/P_{NM}$) for a $1.4 M_\odot$ star,  considering $m_D=5$~GeV (dashed lines) and $m_D=100$~GeV (solid lines) and different values of the DM interaction strength ($y=10^{-1},10,10^3$). The purple dotted line shows the pressure where the hadron-quark phase transition takes place. The dashed-dotted lines indicate unstable configurations.}
\label{fig:PcXratio}
\end{figure}

In order to understand the previously described mass-radius relations,  we should address the quantity of DM necessary for the existence of quark matter in a stable hybrid star for different interaction strengths and DM particle masses. For that purpose, we analyze how the NM central pressure evolves with the accumulation of DM (in terms of the $P_{DM}/P_{NM}$ ratio), paying a special attention when the hadron-quark phase transition occurs. In Fig.~\ref{fig:PcXratio} we display $P_{NM}$ of a $1.4 M_\odot$ star as a function of $P_{DM}/P_{NM}$ for the two DM particle masses cases ($m_D=5$~GeV and $100$~GeV) and different values of the strength interaction $y=10^{-1},10,10^3$. 
We also include a line (purple dotted) representing the pressure at which the hadron-quark phase transition occurs.

%As for $y=10^3$ and $m_D=5$~GeV we get a 1.4M$_\odot$ star only for a few values of $P_{DM}/P_{NM}$ and it is never a hybrid one, we choose, instead, to show the results for intermediary values of $y$ in order to better understand the effect of $y$ for the same $m_D$. 

From Fig.~\ref{fig:PcXratio} we find that the NM central pressure increases with $P_{DM}/P_{NM}$. That, ultimately, is what enables us to get stable hybrid stars. As quark matter starts to appear only at pressures equal to $P_{NM}=361$~MeV/fm$^3$, a star with NM central pressure lower than this value is still completely hadronic, or a hadronic star with DM, but there is no QM in its core. After a certain quantity of DM is added, the central pressure reaches the critical value of 361~MeV/fm$^3$ and the star converts into a hybrid star with a quark core. In the next section, we will discuss the size of this core.

As for the physical meaning of these results, we note the following. When we add DM to NM, the former compresses the latter so that the already dense object gets even more dense so that one needs more NM to get a star with the same mass and approximately the same size. If we go back to Fig.~\ref{fig:NS_branch}, we see that, except for the bottom right plot, we produce a 1.4M$_\odot$ star for different values of $P_{DM}/P_{NM}$ for each DM particle mass and interaction strength, with similar NM radii.
The difference between these $1.4 M_{\odot}$ stars is the quantity of DM and, consequently, of NM. This also explains the breaking of the mass-radius curve after a certain amount of DM is added. Since our NM EoS has an upper pressure limit of 1000 MeV/fm$^3$, the maximum mass it can support decreases as the amount of DM increases. One potential suggestion to address this would be to use an NM EoS capable of reaching higher pressures, allowing for the full mass-radius curve to be obtained. However, several comments are in order regarding this approach. While higher NM pressures would indeed increase the maximum mass reached for the same $P_{DM}/P_{NM}$, there exists a limiting pressure—a threshold beyond which further increases in NM pressure, for the same $P_{DM}/P_{NM}$, no longer raise the maximum mass. At this point, the breaking of the mass-radius curve is governed solely by the amount of DM added, thus the curve will ultimately still stop. This limiting pressure depends on the parameters $y$ and $m_D$ and can be at least two orders of magnitude higher than the upper limit employed in this work.

\begin{figure}[ht]
\begin{centering}
\includegraphics[angle=0,width=0.47\textwidth]{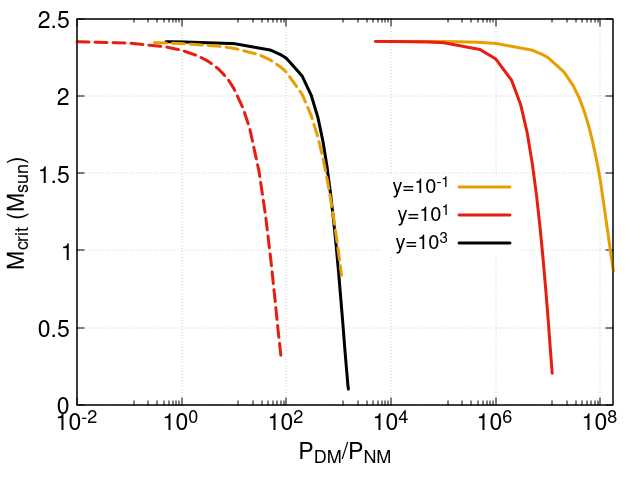}
\caption{Critical mass ($M_{\rm crit}$) as a function of the ratio between pressures ($P_{DM}/P_{NM}$) for $m_D=5$~GeV (dashed lines) and $m_D=100$~GeV (solid lines) and for different values of the interaction strength of the DM ($y$). Only results associated to stable stars are shown. }
%The strongly interacting DM case ($y=10^3$) with $m_D=5$~GeV is left to be analyzed in the last section. }
\label{fig:McXratio}
\end{centering}
\end{figure}

Regarding the change of $M_\text{crit}$ with the accumulation of DM, in Fig.~\ref{fig:McXratio} we plot $M_{\rm crit}$ as a function of $P_{DM}/P_{NM}$ for different values of the DM interaction strength and the two DM masses, as in Fig.~\ref{fig:PcXratio}. We only show results that lead to stable hybrid stars with 
$8\leq R_{NM}\leq 20$~km. As DM compresses NM, the larger the $P_{DM}/P_{NM}$ ratio is,  the larger the NM central pressure becomes, allowing, eventually, for the appearance of QM. In other words, without DM a NM central pressure corresponding to the hadron-quark phase transition pressure, i.e., $P_0=P_C=361$~MeV/fm$^3$ produces a star of 2.35M$_\odot$. However, once enough DM is added, the total mass diminishes so that the same NM central pressure now gives rise to a star with less mass. And, the more DM we add, the less mass this NM central pressure can sustain, so $M_{\rm crit}$ decreases until this star becomes unstable or a DCP (both not shown in Fig.~\ref{fig:McXratio}).

%It is interesting to note that the same critical masses happen almost at the same pressure ratios for $y=10^3$ with $m_D=100$~GeV and for $y=10^{-1}$ with $m_D=5$~GeV.

In Refs.~\cite{lenzi2023dark,Pal_2024} it was also found that the critical mass decreases with the increase of DM. In these works, as the DM interacts with the NM via Higgs boson or scalar and vector meson exchange, 
%even though it does not affect the pressure nor the chemical potential at the hadron-quark phase transition, 
DM makes the EoS softer, which, in turn, leads to a decrease in the maximum mass and the radii move to smaller values.

One could also wonder how the increased central pressure caused by DM influences the appearance and amount of hyperons. The effect is similar to that observed in the hadron-quark phase transition: the presence of DM enables the appearance of hyperons in stars with lower masses. For the NL3$^*\omega \rho$ model used here, the $\Lambda$, $\Sigma^-$, and $\Xi^-$ hyperons emerge in the interior of the compact stars. These hyperons appear when the NM pressure is 44, 64, and 82 MeV/fm$^3$, respectively. These values remain unchanged regardless of the amount of DM added to the star. Therefore, a 1.4M$_\odot$ NS without DM, which has a central pressure of 36~MeV/fm$^3$, will not contain any hyperons. However, once a certain amount of DM is added, which depends on the particle mass and the interaction strength of the DM, as shown in Fig.~\ref{fig:PcXratio}, such a star will contain hyperons.

To close up this section, we summarize our findings. We find that DM does not substitute NM, but that DM allows for more matter to be compressed inside a star, eventually %also 
allowing   QM to appear. The mass and radius of a star are no longer defined exclusively by one central pressure, but by two. It is clear from Fig.~\ref{fig:PcXratio} that various combinations of the two central pressures produces a star with the same mass (1.4M$_\odot$ in that case) and that the addition of DM results in an increase of the NM central pressure of the star. From Fig.~\ref{fig:NS_branch} we see that in most cases even the radius does not change much. Therefore, by only analyzing the mass and the radius of the star, it is impossible to know the NM central pressure and if this central pressure allows the appearance of QM. The presence of DM is then masquerading hybrid stars so that neutron stars without QM in the core and hybrid stars with a sufficient amount of DM have a very similar mass-radius relation. Moreover, hybrid star configurations with DM can be present for such low neutron star masses at which neutron stars with ordinary matter only would not be considered to have a QM core.

%{\color{blue} \bf (Laura: I took the comments in Fig. 4 and 5 of the $y=10^3$ with $m_D=5$~GeV case, since we will analyze it later and we already said that we analyze it later.)}
%\newpage

\subsection{Hadrons, Quarks and Dark Matter in Stars}\label{sec:quantity}

\begin{figure}[ht]
\begin{centering}
\begin{tabular}{cc}
\includegraphics[angle=0,width=0.47\textwidth]{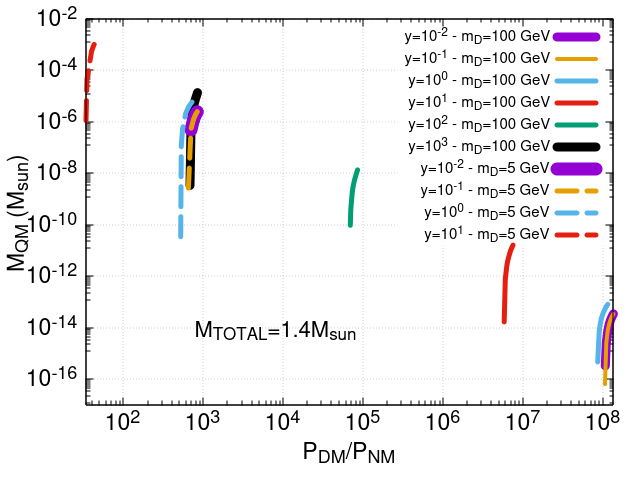} \\%&
\includegraphics[angle=0,width=0.47\textwidth]{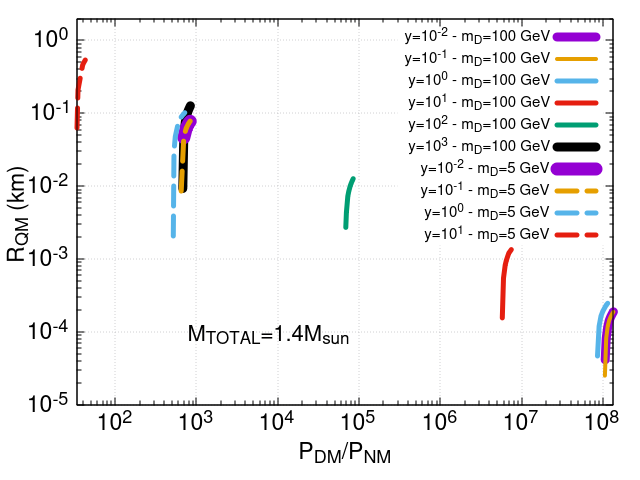} \\
\end{tabular}
\caption{Mass (top) and radius (bottom) of   QM as a function of $P_{DM}/P_{NM}$ in 1.4M$_\odot$ hybrid stars for different interaction strengths $y$ and masses of the DM particle $m_D$.}
\label{fig:M_RqmdmXratio}
\end{centering}
\end{figure}

Now we study the quantity of each type of matter (hadrons, QM and DM) contained in a DM admixed hybrid star. In order to do so we analyze again 1.4M$_\odot$ hybrid stars for different $P_{DM}/P_{NM}$ ratios. In Fig.~\ref{fig:M_RqmdmXratio} we display the total mass of  QM (top panel) and the corresponding radius (bottom panel) as a function of $P_{DM}/P_{NM}$ for stable 1.4M$_\odot$ hybrid stars for $m_D=5$~GeV and $m_D=100$~GeV and different interaction strengths. Note that the results for $m_D=5$~GeV and $y=10^3$ are left for a separate study in Sec.\ref{sec:dark_oysters}.
%Once more we do not include the results for $y=10^2$ and $y=10^3$ with $m_D=5$~GeV because we get only a few stable results for the former and none at all for the latter. 
As can be seen, for each combination of $y$ and $m_D$, the interval of $P_{DM}/P_{NM}$ in which we obtain a stable 1.4M$_\odot$ star that is a hybrid star is very tight. As the total DM masses and the corresponding radii vary even less than the QM ones, we present only the minimum and maximum pressure ratios with the corresponding DM masses and radii in Table~\ref{tab:Mdm_Rdmxratio}.

%As a consequence, the total DM masses and the corresponding radii do not vary much. Because of that, we present only the minimum and maximum pressure ratios with the corresponding DM masses and radii in Table~\ref{tab:Mdm_Rdmxratio}.

\begin{widetext}
\begin{center}
\begin{table}
    \centering
    \begin{tabular}{|c|c|c|c|c|}
    \hline
         $m_D$ (GeV) & $y$ & $P_{DM}/P_{NM}$ & $M_{DM} (M_\odot)$ & $R_{DM} (\text{km})$ \\
        \hline
         $100$ & $10^{-2}$ & $1.07\times 10^8 - 1.35\times 10^8$ & $6.25\times 10^{-5} - 6.15\times 10^{-5}$  & $7.69\times 10^{-4} - 7.05\times 10^{-4}$\\
         \hline
         $100$ & $10^{-1}$ & $1.06\times 10^8 - 1.35\times 10^8$ & $6.26\times 10^{-5} - 6.16\times 10^{-5}$ & $7.72\times 10^{-4} - 7.04\times 10^{-4}$ \\
         \hline
         $100$ & $10^{0}$ & $8.60\times 10^7 - 1.15\times 10^8$ & $6.92\times 10^{-5} - 6.86\times 10^{-5}$ & $8.22\times 10^{-4} - 7.27\times 10^{-4}$\\
         \hline
         $100$ & $10^{1}$ & $5.80\times 10^6 - 7.40\times 10^6$ & $2.57\times 10^{-4} - 2.84\times 10^{-4}$ & $2.57\times 10^{-3} - 2.20\times 10^{-3}$\\
         \hline
         $100$ & $10^{2}$ & $6.90\times 10^4 - 8.50\times 10^4$ & $2.37\times 10^{-3} - 2.66\times 10^{-3}$ & $2.28\times 10^{-2} - 1.97\times 10^{-4}$\\
         \hline
         $100$ & $10^{3}$ & $6.80\times 10^2 - 8.40\times 10^2$ & $2.34\times 10^{-2} - 2.64\times 10^{-2}$ & $2.28\times 10^{-1} - 1.98\times 10^{-1}$\\
         \hline
         $5$ & $10^{-2}$ & $7.00\times 10^2 - 8.40\times 10^2$ & $2.49\times 10^{-2} - 2.45\times 10^{-2}$ & $3.01\times 10^{-1} - 2.81\times 10^{-1}$\\
         \hline
         $5$ & $10^{-1}$ & $6.50\times 10^2 - 8.40\times 10^2$ & $2.50\times 10^{-2} - 2.46\times 10^{-2}$ & $3.09\times 10^{-1} - 2.81\times 10^{-1}$\\
         \hline
         $5$ & $10^{0}$ & $5.20\times 10^2 - 7.20\times 10^2$ & $2.76\times 10^{-2} - 2.73\times 10^{-2}$ & $3.31\times 10^{-1} - 2.89\times 10^{-1}$\\
         \hline
         $5$ & $10^{1}$ & $3.40\times 10^1 - 4.30\times 10^1$ & $9.90\times 10^{-2} - 1.10\times 10^{-1}$ & $1.02 \times 10^{-1} - 8.80\times 10^{-1}$ \\
         \hline
%         $5$ & $10^{2}$ & 1.205 - 1.207 & 4.74\times 10^{-1}$ - 4.75\times 10^{-1}$ & 5.90 - 5.88\\
%         \hline
%         $5$ & $10^{3}$ & never & -  & -\\
%         \hline
    \end{tabular}
    \caption{Minimum and maximum pressure ratios ($P_{DM}/P_{NM}$) that result in stable 1.4M$_\odot$ DM admixed hybrid stars and the corresponding total DM masses ($M_{DM} (M_\odot)$) and radii ($R_{DM} (\text{km})$) for different values of the interaction strength $y$ and DM particle mass $m_D$.%{\color{blue} \bf (Laura: if you do not show $m_D=5$~GeV for strongly interacting DM matter ($y=10^2$ and $y=10^3$) in the plot, do not show them in the table)}
    }
    \label{tab:Mdm_Rdmxratio}
\end{table}
\end{center}
\end{widetext}

The first conclusion we can extract from Fig.~~\ref{fig:M_RqmdmXratio} is that, for a given $m_D$, the larger the DM interaction strength is, the larger the QM masses become.
%and the QM radii are. 
And, for a given DM interaction strength, the smaller the $m_D$ is, the larger the QM masses become.
%and the QM radii become. 
%\bf (Laura: do we have an explanation?)} 
This result can be understood from Fig.~\ref{fig:PcXratio}, where we see that the DM parametrization that is more effective in compressing the star also gives rise to more QM, i.e., the changes in QM follow the ones for DM. In \cite{narain2006compact,barbat2024comprehensive} it was shown that the total mass of a DM fermionic star grows as the particle mass decreases ($M \propto m_D^{-2}$) and as the interacting strength increases. And in Fig.~~\ref{fig:M_RqmdmXratio} we can see the same behavior for the QM core, as QM feels the gravitational potential generated by DM, that varies with the interacting strength and/or DM particle mass.

Moreover we find that for each value of $m_D$ and $y$
the larger the $P_{DM}/P_{NM}$ ratio is, the more   QM is produced. This can be easily understood if we look again at Fig.~\ref{fig:PcXratio}, the increase of the $P_{DM}/P_{NM}$ ratio also induces larger NM central pressure, which, of course, results in more   QM. As for DM, the quantity of DM  does not vary much as the interval for $P_{DM}/P_{NM}$ for which we obtain stable stars is tight, remaining almost constant, as seen in 
Table~\ref{tab:Mdm_Rdmxratio}. In fact, from Table \ref{tab:Mdm_Rdmxratio} we observe that the DM contribution to the mass slightly decreases when augmenting the pressure ratio for $10^{-2} \leq y \leq 1$, whereas it increases for higher values of $y$ for both values of $m_D$.  In \cite{narain2006compact} it is shown for a single DM fermionic star that, when keeping the same central pressure and changing the interacting strength, for small values of $y$, the maximum mass does not change, while it increases with a power law for strong interactions ($y>1$). Here the DM central pressure is increased, but for small values of $y$ the DM looses the 'competition' to NM, which then appears in the form of   QM. So, for small values of $y$, when increasing the $P_{DM}/P_{NM}$, the contribution to the total mass from QM increases and from DM decrease, whereas for $y>1$ the contributions from both types of matter increase with the pressure.

If we now compare the amount of   QM mass with respect to the DM mass, we find that the contribution to the total mass coming from DM is much larger. In spite of this, 
%, the most DM mass contribution we can get, which happens when $y=10^2$, $m_D=5$~GeV and $P_{DM}/P_{NM}=1.207$, corresponds to less than 35$\%$ of the total mass, showing that 
the maximum contribution to the total mass from DM corresponds to less than 8$\%$ of the total mass (for $y=10$, $m_D=5$~GeV and $P_{DM}/P_{NM}=43$), showing that
the most important contribution to the total mass still comes from hadronic matter. %{\color{blue} \bf (Laura: I took out the example of $y=10^2$, $m_D=5$~GeV because this is part of the special case in the next section. Could you put some numbers for another $y$ and $m_D$?)}
%For the same values of $y$, $m_D$ and $P_{DM}/P_{NM}$ we also get the most   QM, which is $M_{QM}=$1.95\times 10^{-3}$M$_\odot$, that corresponds to only 0.14$\%$ of the total mass of the star.
For the same values of $y$, $m_D$ and $P_{DM}/P_{NM}$ we also get the most QM, which is $M_{QM}=9.85\times 10^{-4}M_\odot$, that corresponds to only 0.07$\%$ of the total mass of the star.

%%For example, for $y=10^{-2}$ and $m_D=100$~GeV, the DM configuration to which we get the least QM and DM, we can get a mass up to $M_{QM}=$3.38\times 10^{-14}$M$_\odot$ (or 6.72\times 10^{16}$ kg) whereas for the DM the maximum mass we get is $m_D=$6.25\times 10^{-5}$M$_\odot$ (or 1.24\times 10^{26}$ kg). And for the $y=10^{2}$ and $m_D=5$~GeV, not included in the plots, we get $M_{QM}=$1.95\times 10^{-3}$M$_\odot$ (or 3.88\times 10^{27}$ kg) and $m_D=$4.75\times 10^{-1}$M$_\odot$ (or 9.45\times 10^{29}$ kg). So, it is clear that the major contribution to the total mass comes from the hadronic matter. 

With regard to the radii for each type of matter, we obtain that QM accumulates within a small radius at the core of the star, while following the trend of the QM mass, that is, smaller QM masses also mean smaller QM radii, with the radius increasing as the mass augments.  The QM radii can be as small as  4.14~cm and reach at most 0.539~km for $y=10$ and $m_D=5$~GeV.
%\st{(for $y=10^2$ and $m_D=5$~GeV, not shown in the plot, we get a radius of 0.749 km)}. 
As for the DM radii, for all the cases shown here, its value is also small but slightly bigger than the QM radii, with values varying from  70.4~cm up to 1.02~km. Note that the DM radius is also quite insensitive to $P_{DM}/P_{NM}$, slightly decreasing with the ratio, as the increase of the ratio compresses DM.
% {\color{blue} \bf (Laura: do we have an explanation? } 
So, in conclusion, most of DM is admixed with   QM in the most inner core of 1.4M$_{\odot}$, but a small amount of DM is also mixed with hadronic matter.
%{\color{blue} This also explains why there is more DM mass than QM in the star.}
%(Laura: why? Carline: because the DM has more space - and also, DM is allowed at all pressure values and QM only at $P_{NM}>361$) Laura: My question is "This also explains.." What explains what? I think you are just repeating your conclusion and not given a further argument. Or are you?}

%{\color{blue} \bf (Laura: I have the feeling we are just describing without giving any explanation. We have to think a bit more)}
%Its value is also not as sensitive to the pressure ratios% as the QM radius
%, but here, in all cases, %for the same combination of $y$ and $m_D$ 
%the radius decreases with the increase of the pressure ratio. 

Up to now we have shown the composition and structure of 1.4M$_\odot$ DM admixed hybrid stars. However, one may pose the question of the interplay between QM and DM for larger star masses. Stars with larger masses have higher central pressures without DM. Hence, the ratio $P_{DM}/P_{NM}$  needed to reach the hadron-quark phase transition pressure is smaller and, as a direct consequence, the quantity of DM in those stars will be smaller. However, the amount of QM can be larger. On the other hand, stars with less than 1.4M$_\odot$ will require more DM as compared to those with larger masses so as to reach the hadron-quark phase transition. Therefore, these low-mass stars will have a bigger and more massive DM core, but a less massive QM core. At the same time, the radius of the DM decreases, so that the DM core of these stars is denser. %But, if we look back at Fig.~\ref{fig:McXratio} we see that, depending on the type of DM, there is a limit in the minimum mass of the star that can be hybrid. That is because, eventually, the DM renders the low mass stars unstable.

Note that in Ref.~\cite{lenzi2023dark} it was also found that the QM core increases with the increase of DM content. The authors showed that, for example, a QM core of a 2M$_\odot$ star that already corresponds to approximately 60$\%$ of the mass of the star when there is no DM present, can grow up to 80$\%$ of the total mass when DM is added.
 %(Laura: is not the opposite of what we are saying in the previous paragraph? In that paragraph we were saying that stars with masses of above 1.4 Msun, have larger QM masses and smaller DM masses??? Carline: I am not sure if I understand the problem. In the previous paragraph I wanted to say that stars with masses above 1.4 Msun have larger QM masses and smaller DM masses than 1.4 Msun. Here I am merely saying that ~\cite{lenzi2023dark} also found that the addition of DM causes the increase of the QM core - maybe u understood that the mass of the star is growing? I wanted to say that the mass of the QM core is growing. I rephrased it a bit and hope that it is clearer now)}

It is also worth mentioning that we obtain very small QM cores as compared to the ones obtained in other works that explore the possible existence of hybrid stars without DM \cite{lopes2022hypermassive} or in the one-fluid model of interacting NM with DM, as done in Ref.~\cite{lenzi2023dark}. However, we remind the reader that our present goal is to show that DM can trigger the appearance of QM. Thus, we choose hadronic and QM parametrizations that do not produce stable hybrid stars without DM. With the conclusions drawn here, we can argue that for NM configurations that already allow the appearance of   QM, with the addition of DM, the amount of QM can be even larger.

%{\color{blue} \bf (Laura: you might want to comment on Ref. [56] where Debora and others use the same NM parametrization but coupling DM via the Higgs portal to obtain DM admixed hybrid stars. And \cite{Pal_2024}. Maybe others?)}

%for one y and m_D, when I am increasing the pressure ratio, when I am looking at the quantity of QM mass, the NM pressure is always around 360 (I integrate from the maximum pressure until the phase transition pressure - I am only looking at the 1.4M_sun star). So, as the pressure ratio increases, that means that the DM pressure has to diminish. - maybe include this in the discussion above

%to get the DM contribution to the mass I integrate from the maximum pressure (to generate a 1.4M_sun) until I reach 'zero' for both pressures

\subsection{Dark Oysters}\label{sec:dark_oysters}

In this last section, our aim is to analyze
the case of strongly interacting DM ($y=10^3$) with a particle mass of $m_D=5$~GeV, given the particular mass-radius configurations obtained in Fig.~\ref{fig:NS_branch}.
 
In the bottom right plot of Fig.~\ref{fig:NS_branch} we show that the mass-radius configurations are much more sensitive to changes in the $P_{DM}/P_{NM}$ ratio,
%to this parametrization of DM, already changing at very low $P_{DM}/P_{NM}$ and,
as compared to the other three cases depicted in the same figure. Also, the mass of the DM admixed %hybrid 
stars increases with the pressure ratio. These results stem from the fact that DM dominates over NM as we increase the ratio, so that the total mass is governed by the mass of the DM component, which scales with the inverse of the square of the DM particle mass, as already discussed in ~\cite{narain2006compact,dengler2022second}. In fact, already at $P_{DM}/P_{NM}=10^{-1}$ the DM mass starts to exceed the NM mass and at $P_{DM}/P_{NM}=10^3$ the DM completely dominates so that a further increase of the ratio will not change much the shape of the M$_T-R_{DM}$ relation, where M$_T$ is the total mass, even though the M$_T- R_{NM}$ relation will still slightly change, as NM will be further compressed into smaller radii.

\begin{figure}[ht]
\begin{centering}
\begin{tabular}{cc}
\includegraphics[angle=0,width=0.47\textwidth]{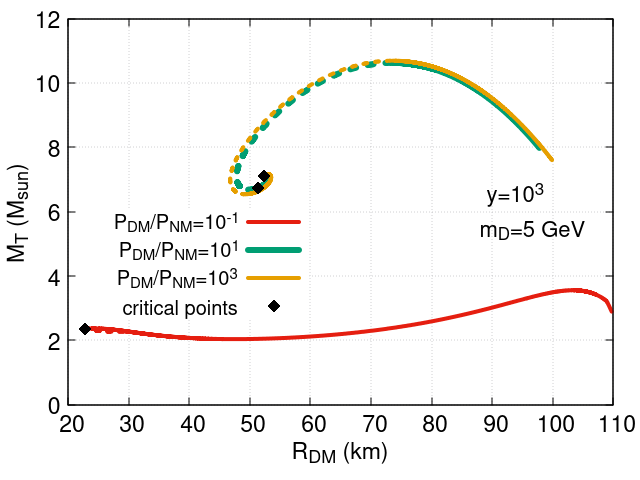} \\
\includegraphics[angle=0,width=0.47\textwidth]{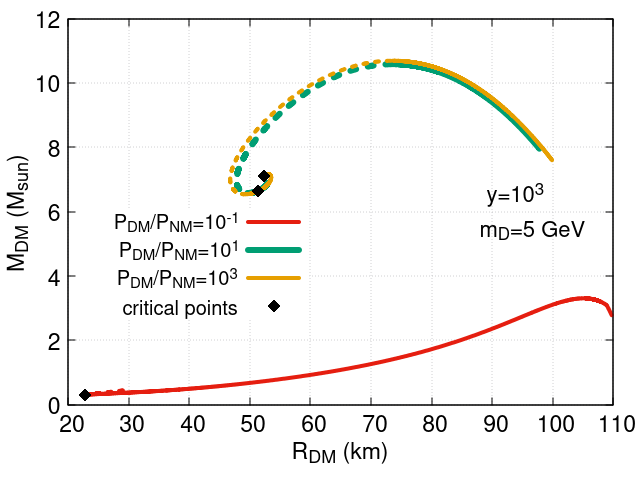} \\
\end{tabular}
\caption{Total mass (top panel) and DM mass (bottom panel) as a function of the DM radius for strongly  interacting DM ($y=10^{3}$) and $m_D=5$~GeV for different pressure ratios for values of the observable radius $R_{NM}<20$~km. The dots indicate the critical mass $M_{\rm crit}$ at which hybrid stars start to appear for each configuration. We also include the unstable results (dashed lines). Note that $M_{\rm crit}$ for $P_{DM}/P_{NM}=10^3$ is actually also unstable.}
\label{fig:NS_branch-DM}
\end{centering}
\end{figure}

In order to show the dominance of DM over NM, in Fig.~\ref{fig:NS_branch-DM} we display the total mass (top panel) and only the DM contribution to the total mass (bottom panel) as a function of the DM radius for the same $P_{DM}/P_{NM}$ shown in the bottom right plot of Fig.~\ref{fig:NS_branch}. We consider DM admixed %hybrid 
stars with the normal matter radius $R_{NM}<20$~km. First, we observe that the DM radius is always much larger than the observable one ($R_{NM}$). In Table~\ref{tab:dark_oyster} we show the features of some of the stable objects displayed in Fig.~\ref{fig:NS_branch-DM}. 

%Given that we obtain stars with masses larger than 1.4M$_\odot$, we take the smallest star displayed in Fig.~\ref{fig:NS_branch-DM}, that is, the one with the smallest DM radius for $P_{DM}/P_{NM}=10^{-1}$. In this case we have an object whose features are given in Table~\ref{tab:dark_oyster}.

\begin{widetext}
\begin{center}
\begin{table}
    \centering
    \begin{tabular}{|c|c|c|c|c|c|c|}
    \hline
        $P_{DM}/P_{NM}$ & M$_{T}$ (M$_\odot$) & M$_{NM}$ (M$_\odot$) & M$_{DM}$ (M$_\odot$) & R$_{NM}$ (km) & R$_{DM}$ (km) & $P_{NM}$ (MeV/fm$^3$) \\
        \hline
        $10^{-1}$ & 2.36 & 2.05 & 0.31 & 11.9 & 22.8 & 362\\
         \hline
         $10^{-1}$ & 2.12 & 1.19 & 0.93 & 12.9 & 60.0 & 33 \\
         \hline
         $10^{-1}$ & 2.90 & 0.11 & 2.78 & 20.0 & 109.7 & 1 \\
         \hline
         $10^{1}$ & 6.95 & 6.36 $\times$ 10$^{-2}$ & 6.88 & 3.2 & 53.0 & 997 \\
         \hline
         $10^{1}$ & 6.68 & 0.10 & 6.58 & 4.60 & 49.78 & 199 \\
         \hline
         $10^{1}$ & 7.98 & 2.59 $\times$ 10$^{-2}$ & 7.96 & 20.0 & 97.8 & 8 $\times$ 10$^{-2}$ \\
         \hline
         $10^{3}$ & 7.16 & 1.72 $\times$ 10$^{-4}$ & 7.16 & 0.60 & 53.0 & 75.0 \\
         \hline
         $10^{3}$ & 6.54  & 2.53 $\times$ 10$^{-4}$ & 6.54 & 1.24 & 49.3 & 2 \\
         \hline
         $10^{3}$ & 7.60 & 9.12 $\times$ 10$^{-4}$ & 7.60 & 20.0 & 99.9 & 6 $\times$ 10$^{-4}$ \\
         \hline
    \end{tabular}
    \caption{Masses and radii of the normal matter and dark matter components for some of the stable objects obtained for $P_{DM}/P_{NM}=10^{-1}, 10^{1}$ and $10^{3}$  with $y=10^3$ and $m_D=5$~GeV. We also show the value of the central pressure for normal matter. Note that only for the first two stars there is still a dominance of NM over DM, i.e., with $M_{DM}/M_{NM}<1$.}
    \label{tab:dark_oyster}
\end{table}
\end{center}
\end{widetext}

%\caption{Masses and radii of the hadronic, and dark components for a 2.36M$_\odot$ star, obtained for $P_{DM}/P_{NM}=10^{-1}$ with $y=10^3$ and $m_D=5$~GeV.} 

As seen in Table~\ref{tab:dark_oyster}, all of these compact objects have a core formed by a mixture of NM and DM, surrounded by a DM halo. The structure of these objects is 
%this compact object has a core of a radius of 11.9 km formed by a mixture of NM and DM, surrounded by DM up to 22.8 km. The structure of this star is 
completely different to what we obtained for the other DM parametrizations explored in the previous sections ($m_D=100$~GeV as well as $m_D=5$~GeV for the weakly interacting case), where the DM radius was always much smaller than the NM one, so that the DM accumulated only in the core of the stars. Actually, these mass-radius configurations for low $m_D$ and strongly interacting DM were reported in Refs.~\cite{deliyergiyev2019dark,dengler2022second,barbat2024comprehensive}, but here we find an even more extreme scenario, with the DM radius exceeding more than four times the size of the core in all cases but one. Because these configurations resemble oysters, with the DM halo representing a large black shell and the NM core the small bright shining pearl (even though there is also DM mixed in this core), we have named these objects dark oysters.

\begin{figure}[ht]
\includegraphics[angle=0,width=0.47\textwidth]{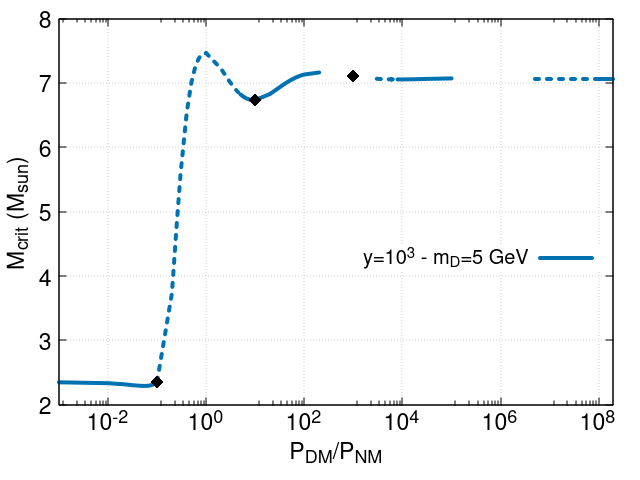} 
\caption{Critical mass $M_{\rm crit}$ as a  function of the DM pressure ratios $P_{DM}/P_{NM}$ for strongly interacting DM $y=10^3$ and particle mass $m_D=5$~GeV. The solid lines represent stable stars. As for the dashed lines, the first hybrid star is unstable, but, at higher central pressures, maintaining the same ratio $P_{DM}/P_{NM}$, stable ones appear. The dots correspond to the critical masses for the three cases shown in Fig.~\ref{fig:NS_branch-DM}.} 
\label{fig:McritXratio-DO}
\end{figure}

%And what about the   QM? 
As DM is dominating and increasing the mass of the stars, one would expect that the critical mass would also augment. And, in fact, that is what we observe when we plot the critical mass $M_{\rm crit}$ as a function of $P_{DM}/P_{NM}$. As can be seen in Fig.~\ref{fig:McritXratio-DO}, the behavior of the critical mass is very different from what we have seen in the previous sections. For low $P_{DM}/P_{NM}$, $M_{\rm crit}$ is  decreasing with the increase of the $P_{DM}/P_{NM}$. However, at $P_{DM}/P_{NM}=8 \times 10^{-2}$ the behavior completely changes and $M_\text{crit}$ quickly increases. %At this pressure ratio the critical mass also turns unstable.

Apart from the different behavior of the critical mass with the $P_{DM}/P_{NM}$ ratio, we find that some values of $M_{\rm crit}$ correspond to unstable stars.
In Fig.~\ref{fig:McritXratio-DO} the dashed lines represent these unstable solutions, where the star with $P_{NM}=361$~MeV/fm$^3$, which is the pressure where the hadron-quark phase transition happens, is unstable.  However, 
as the NM central pressure is increased, for the same $P_{DM}/P_{NM}$, we get stable stars again, which, of course, are hybrid. This would be the case of $P_{DM}/P_{NM}=10^1$ or $P_{DM}/P_{NM}=10^3$ if $M_{\rm crit}$ were located somewhere between $50<R_{DM}<70$~km in Fig.~\ref{fig:NS_branch-DM} or $5<R_{NM}<9$~km in the bottom right of Fig.~\ref{fig:NS_branch}.
%(Laura: I am confused, where do I see this? Carline: I tried to plot an example using a ratio between 10^{-1}$ and 10^{1}$, but it would always mess up something else I wanted to show. But it would be, for example, if for $10^1$ or $10^3$ the critical mass would fall between DM radius 50 and 70 (dashed lines in fig 7))} 
The solid lines represent stable solutions and the dots correspond to the critical masses shown in Fig.~\ref{fig:NS_branch-DM}. Note that the critical mass for $P_{DM}/P_{NM}=10^3$ falls into a gap in the $M_{\rm crit} \times P_{DM}/P_{NM}$ line. This is because this $M_{\rm crit}$ corresponds to unstable star configurations and, differently to the results that fall on the dashed lines, here, with the increase of the central pressure no stable hybrid star ever appears in our calculations.
%\footnote{This is hard to see from Fig.~\ref{fig:NS_branch-DM} and from bottom right of Fig.~\ref{fig:NS_branch} because all results for $P_{DM}/P_{NM}=10^3$ for stars with $P_{NM}\gtrsim 300$ MeV/fm$^3$ are very similar.}

%, {\color{blue} as seen in Fig.~\ref{fig:NS_branch-DM} for $P_{DM}/P_{OM}=10^{-1}, 10$???}.
%The solid lines in Fig.~\ref{fig:McritXratio-DO}   represent the stable solutions. {\color{blue} \bf (Laura: this is difficult to see in Fig.8 only, should we not refer to Fig. 7? I think we have to discuss Fig. 8 with Fig. 7 to understand what you are saying) }

At this point we should mention that, although the stability analysis used here is a step forward from the naive analysis done in previous works as we consider here the changes that the NM may induce on the DM and vice versa, it is not totally clear that, at high pressures, the change in stability showed by our stability analysis really corresponds to the change of the lowest energy mode $\omega_0$. So, our results for large pressures should be regarded with a grain of salt. A rigorous stability analysis of two-fluid stars is left to future work.

Regarding the quantity of QM, it can be larger than any of the cases analyzed in the previous section, which might be counterintuitive, considering that DM is strongly dominating over the NM. As we can see from Fig.~\ref{fig:NS_branch-DM}, for $P_{DM}/P_{NM}=10$ the contribution to the total mass coming from the NM seems negligible. In fact, for the star with $M_T=6.95$M$_\odot$, NM only contributes with $6.36 \times 10^{-2}$M$_\odot$. However, for the same star,   QM accounts for $2.01 \times 10^{-2}$M$_\odot$, which is more than we did obtain in the previous cases (see Fig.~\ref{fig:M_RqmdmXratio}). Of course, the relative contribution of QM to the total mass is much less than in the previous cases.

%The nature of the DM, i.e., its interaction strength and the mass of its particle, and the fact that we choose to analyse only results with a observable radius smaller than 20 km is what allowed us to reach the strong dominance of the DM over the NM here but not for the cases analyzed in the previous sections.

Before finalizing this section, we should make some comments about the dominance of DM. We have analyzed stars with $R<20$km for $y=10^3$ with $m_D=5$~GeV, where DM dominates. In this case, the effect of DM is to increase the total mass of the stars, and, at the same time to decrease the observable radius. For the other combinations of $y$ and $m_D$ explored in the previous section, we have concentrated on the NS branch and DM that is not the dominant component. In that context, the effect of the DM is to first 'eat' away the NS branch, making the results migrate to the white dwarf branch, i.e., to larger radii (larger than 20 km) and then, eventually, to smaller masses and radii to form the DCPs. This means that, if we keep on increasing the pressure ratios for that cases so that the DM becomes dominant, eventually we would obtain very small $M_\text{crit}$ associated to objects with very small radii, that is, we would observe hybrid dark compact planets. The exploration of such results is kept for future works. 

As already mentioned above, the increase in the total mass of DM admixed stars at low $m_D$ was already reported in~\cite{deliyergiyev2019dark,dengler2022second,barbat2024comprehensive}, although the discussion on the DM halo was scarce. In fact, those works show that the dark oyster configurations are also possible for weakly interacting DM as long as the particle mass is decreased even further. No work, however, to the best of our knowledge, has ever reported on dark oysters with a hybrid core.

%{\color{blue} \bf (Laura: so where is the dark oyster definition? You had written ¨dark oyster"  in caption of Table V but there was no explanation, so I remove it)}

%{\color{blue} \bf (Laura: comparison with other works?)}

%{\color{red} should I include something about $y=10^2$ here? I also, eventually, get a dark oyster...maybe just mention it - at lower pressure ratios I get very few hybrid stars with more QM than for the cases showed in the previous section, but if I increase the ratio at some point I get dark oysters}

\section{Conclusions}

In this paper, we have investigated the effect of DM on hybrid stars using a two-fluid approach, considering that NM and DM interact only gravitationally. 
%This approach distinguishes our work from the only other two studies in the literature that have explored this subject, namely \cite{lenzi2023dark} and \cite{Pal_2024}. 
For NM we have built an EoS for hybrid stars via the Maxwell construction from a QHD-based model, for nucleons and hyperons, and a MIT-based model for uds-quark matter.  For the DM EoS we use a non-selfannihilating self-interacting Fermi gas with different interaction strengths and two different
particle masses, $m_D=5$~GeV and $m_D=100$~GeV.

We have found that the presence of DM in NSs may trigger the appearance of QM in its core at unprecedented low masses, giving rise to DM-admixed hybrid stars. This happens because the presence of DM causes an increase of the central pressure of the NS, which, after a certain amount of DM is added, reaches the hadron-quark phase transition value. In a star that already has a QM core, the presence of DM will enhance this core. The amount of DM (in terms of the $P_{DM}/P_{NM}$ ratio) that needs to be accumulated in order for the central pressure to increase depends on its interaction strength $y$ and particle mass $m_D$: for weakly interacting DM one needs to add more DM than for the strongly interacting case and for the same interaction strength $y$, one needs to add more DM when the DM particle mass is larger. A direct consequence of this result is that the critical mass, i.e., the minimal mass of the star with a quark core, decreases with the increase of $P_{DM}/P_{NM}$. Except for strongly interacting DM with $m_D=5$~GeV, we have shown that, by only analyzing the mass and the radius of the star, it is impossible to know the NM central pressure and if this central pressure allows the appearance of QM. The DM is then masquerading hybrid stars. 

We have also determined that the quantity of each type of matter (hadron, quark and dark) that can accumulate in such stars depends on the  interaction strength and DM particle mass, so that the DM parametrization that is more effective in compressing the star also gives rise to more   QM, i.e., the changes in QM follow the ones for DM. Except for the strongly interacting DM with $m_D=5$~GeV, we have obtained DM-admixed hybrid stars with very small QM cores and slightly larger DM cores.

In the last section we have carefully analyzed the results for a DM where $y=10^3$ and $m_D=5$~GeV. In this case we obtained what we named dark oysters, stars with a large DM radius and small observable radius of ordinary matter. 
Here, the addition of DM augments the total mass of the star. Such results where already observed in Refs.~\cite{deliyergiyev2019dark,dengler2022second,barbat2024comprehensive}, but to the best of our knowledge, the present work is the first to report on dark oysters with a hybrid core. In fact, for the dark oyster we have shown that QM mass can be larger than for the other DM parametrizations we have explored in this work. Dark oysters could be observable by measuring, for example, neutron stars with seemingly unphysically small radii of just a few kilometres, while having large total gravitational masses of several solar masses.

%{\color{blue}can we say something about possible ways to detect DM admixed hybrid stars? }

\section*{Acknowledgements}

C.B. received support from the Conselho Nacional de Desenvolvimento Científico e Tecnológico (CNPq/Brazil) and thanks the Institute of Space Sciences (ICE, CSIC) for hosting her during a year.
L.T. acknowledges support from CEX2020-001058-M (Unidad de Excelencia ``Mar\'{\i}a de Maeztu") and PID2022-139427NB-I00 financed by the Spanish MCIN/AEI/10.13039/501100011033/FEDER,UE as well as from the Generalitat de Catalunya under contract 2021 SGR 171 and from the Generalitat Valenciana under contract CIPROM/2023/59. 
L.T. and J.S.B. acknowledge support by the CRC-TR 211 'Strong-interaction matter under extreme conditions'- project Nr. 315477589 - TRR 211. 

%\newpage

%\clearpage

%\begin{center}
   \section*{References} 
%\end{center}

\bibliography{references}

%\bibliographystyle{acm}

%\begin{thebibliography}{}

%\end{thebibliography}

\end{document}